\documentclass[10pt]{article}
\usepackage{graphicx}
\usepackage{appendix}
\usepackage{verbatim}
\usepackage{amsmath}
\usepackage{amsfonts}
\def\al{\alpha}
\def\be{\beta}
\def\ga{\gamma}
\def\de{\delta}
\def\dal{{\dot\alpha}}
\def\dbe{{\dot\beta}}

\def\eps{\varepsilon}
\def\vp{\varphi}
\def\th{\theta}
\def\ka{\kappa}
\def\la{\lambda}
\def\si{\sigma}
\def\bsi{\bar{\sigma}}
\def\om{\omega}

\def\hc{{\rm h.c.}}
\def\cA{{\mathcal A}}
\def\cC{{\mathcal C}}
\def\cD{{\mathcal D}}
\def\cJ{{\mathcal J}}
\def\cL{{\mathcal L}}
\def\cK{{\mathcal K}}
\def\cM{{\mathcal M}}

\def\cO{{\mathcal O}}
\def\cP{{\mathcal P}}
\def\cR{{\mathcal R}}

\def\cU{{\mathcal U}}
\def\11{{\mathbb 1}}

\def\JJ{{\mathbb J}}
\def\MM{{\mathbb M}}

\def\rd{{\rm d}}
\def\re{{\rm e}}
\def\ri{{\rm i}}

\def\JJBL{\JJ_{B\!-\!L}}
\def\cJBL{\cJ_{B\!-\!L}}

\def\rX{{\rm X}}
\def\hrX{\widehat{\rm X}}

\def\tF{\widetilde{F}}
\def\tG{\widetilde{G}}

\def\bel{\bar{e}}
\def\bd{\bar{d}}
\def\bu{\bar{u}}
\def\bN{\bar{N}}
\def\bnu{\bar{\nu}}

\def\mf{\mathfrak{m}}

\def\beq{\begin{equation}}
\def\eeq{\end{equation}}
\def\bea{\begin{eqnarray}}
\def\eea{\end{eqnarray}}
\def\btp{\begin{tikzpicture}}
\def\etp{\end{tikzpicture}}
\def\nn{\nonumber}

\def\rarr{\rightarrow}

\def\lrarr{\leftrightarrow}

\def\dslash{\partial \hspace{-1.3ex}\slash}
\def\dslashbar{\bar\partial \hspace{-1.3ex}\slash}
\def\pslash{p \hspace{-1.0ex}\slash}
\def\pslashbar{\bar{p} \hspace{-1.0ex}\slash}
\def\PS{p\hspace{-1.0ex}\slash}
\def\PSB{\bar{p} \hspace{-1.0ex}\slash}

\def\kslash{k \hspace{-1.0ex}\slash}

\def\qslash{q \hspace{-1.3ex}\slash}

\def\me{m_{e_i}}
\def\md{m_{d_K}}
\def\vev{\langle\varphi\rangle}

\begin{document}

\begin{center}
{\bf \Large Neutrino Mixing and the Axion-Gluon Vertex}\\[1.5cm]

{\bf Adam Latosi{\'n}ski$^1$, Krzysztof A. Meissner$^{1,2}$ and Hermann Nicolai$^2$}

\vspace{1cm}
{{\it $^1$ Faculty of Physics,
University of Warsaw,
Ho\.za 69, Warsaw, Poland\\
$^2$ Max-Planck-Institut f\"ur Gravitationsphysik
(Albert-Einstein-Institut)\\
M\"uhlenberg 1, D-14476 Potsdam, Germany\\
}}
\end{center}

\vspace{2cm}

{\footnotesize
{We present detailed arguments and calculations in support of our recent
proposal to identify the axion arising in the solution of the strong CP problem
with the Majoron, the (pseudo-)Goldstone boson of spontaneously broken lepton
number symmetry. At low energies, the associated $U(1)_L$ becomes, via
electroweak parity violation and neutrino mediation, indistinguishable
from an axial Peccei-Quinn symmetry in relation to the strong interactions.
The axionic couplings are then fully computable in terms of known  SM parameters
and the Majorana mass scales. The determination of these couplings involves certain three-loop diagrams, with a UV finite neutrino triangle taking over the role of the usual triangle
anomaly. A main novelty of our proposal is thus to solve the strong CP problem
by a non-local term that produces an anomaly-like term in the IR limit.}
}

\newpage

\section{Introduction}

In this paper we provide detailed arguments and calculations for our recent
proposal to identify the axion with the Majoron \cite{MNaxion}. This proposal
rests on very special features of the neutrinos and their mixing which, to the best of our knowledge, have not been fully exhibited in the existing literature. The present work grew
out of an earlier attempt \cite{MN} to incorporate classical conformal symmetry into the Standard Model (henceforth abbreviated as `SM'), and thereby resolve some
of the outstanding issues of SM  physics in a minimalistic fashion, that is,
{\em without} introducing any large intermediate mass scales or unobserved new particles beyond the known SM spectrum other than right-chiral neutrinos and one new complex
scalar field (and, in particular, no low energy supersymmetry).\footnote{For
 alternative   proposals in this direction, see
 \cite{Bardeen,Hempfling,Canadians,Shap,Shap1,Foot,Ito,Lindner,Pilaftsis,Dias} and
 references therein. In particular, our model is similar  to the so-called $\nu$MSM first
 introduced in \cite{nuMSM}  and further studied in \cite{Shap,Shap1}, except
 that the extra scalar field $\phi(x)$ here is complex and {\em not}
 supposed to be identified with the inflaton.}
Any such proposal must in particular account for the two phenomena whose explanation is commonly attributed to new scales intermediate between the electroweak scale
and the Planck scale, namely {\em light neutrino masses} and {\em axion couplings}.
In fact, it was understood already some time ago \cite{Akhmedov} that the first item can be
achieved by taking the neutrino Yukawa couplings to be $\sim \cO(10^{-5})$ (an
acceptable fine-tuning in view of the fact that the mass ratio $m_u/m_t$ is of the same order of magnitude). The present work, then, addresses the question whether one can likewise
solve the strong CP problem {\em without} new mass scales of order
$ > 10^{10} \, {\rm GeV}$. Consequently, the model that we will study in this paper
is {\em a minimal extension of the SM with only right-chiral  neutrinos and one new
complex scalar field $\phi(x)$}, but {\em without} explicit mass terms for the fermions (and in particular, no explicit Majorana mass). For definiteness we will refer to it as the {\em Conformal Standard Model} (abbreviated as `CSM') in the remainder.

Apart from underlining the potential importance of conformal symmetry in the SM,
the present proposal furnishes an (in our view) intriguing link between strong
interaction physics, where the axion  is needed to solve the strong CP problem
\cite{PQ,WW,VW,DiV}, and the electroweak sector of the SM, where the Majoron arises as the Goldstone boson of spontaneously broken lepton number symmetry \cite{Mo1,Mo3}
(a possible link between axions and neutrinos had already been suggested in
\cite{LPY,Khlopov1,Khlopov2}).
This link is encapsulated in the relation \cite{MNaxion}
\beq\label{faN}
f_a^{-1} \,\propto \, \frac{\al_w^2}{\cM^2} \, \sum m_\nu
\eeq
tying the axion coupling $f_a$ to the ratio of the sum of the light neutrino masses and a
(mass)$^2$ parameter $\cM^2$, where $\cM$ is of order $M_W$ or the largest of the heavy neutrino mass eigenvalues; $\al_w$ is the $SU(2)_w$ gauge coupling (the estimate (\ref{faN})
is based on the formula (\ref{faxion}), which is obtained from a three-loop
calculation and  more complicated, see the final section of this paper). The weakness
of axionic couplings can thus be naturally understood and explained,
without adducing any large intermediate scale.

The main new features of our proposal can be summarized as follows:
\begin{itemize}
\item
 If the axion is identified with the Majoron, all its couplings  arise as effective couplings via loop corrections with the insertion of a triangle diagram involving neutrinos on the internal lines. There is consequently no need for new global symmetries beyond the global symmetries already present in the CSM (baryon and lepton number symmetry).
\item  This triangle diagram, though superficially reminiscent of the well known anomalous triangle diagrams in QED and QCD, is UV finite and produces a
 {\em non-local} contribution to the effective action, which however reduces to the (local) anomaly-like amplitude in the low momentum limit. Hence,
 a main novelty of our proposal is to solve the strong CP problem via a non-local term that coincides with the usual anomaly only in the IR limit.
\item
 The non-vanishing result for the amplitude hinges on the spontaneous breaking
 of lepton number symmetry (as well as on electroweak symmetry breaking), and on the fact that parity is maximally violated in the SM. If lepton number symmetry were restored, the relevant diagrams would vanish identically even if electroweak symmetry remained broken.
\item
 If spontaneous symmetry breaking and the emergence of mass scales can be
  accounted for by a Coleman-Weinberg-type mechanism, starting from a conformal
  classical Lagrangian, all effects can be viewed as ultimately resulting from
  the associated conformal anomaly.
\end{itemize}

We here concentrate on the link with strong interaction physics, while the
discussion of photon couplings, as well as implications for cosmology,
especially with the axion as a Dark Matter candidate,
will be analyzed elsewhere. However, before entering into any
details of the calculation it is perhaps worthwhile to clearly state what the result provided at the end of this paper actually means. We want to calculate the effective coupling of the axion to
two gluons in lowest non-vanishing loop order. The term `effective' in this context
means that in the effective lagrangian we leave only massless or almost massless
particles, namely axions, light neutrinos, gluons and photons (but the latter will not be
considered in this paper, as we said), and therefore `integrate out'  all massive
fields, whose masses are much larger than the external momenta, and which
do not appear on the external legs  (i.e. charged leptons and right-chiral
neutrinos, quarks, massive gauge bosons,  and all scalar fields).
As we will see, there are many diagrams that should be considered  at one
and two loops, but it turns out that almost all of them give vanishing contribution
to the coupling at hand because of the powerful fact that all
gauge anomalies cancel in the SM. Therefore only the three-loop diagrams studied
in the last section actually contribute to the effective coupling.
Nevertheless, we find it convenient to divide the calculation into three steps
associated to the different loop orders, but it is important to keep in mind that the results of these intermediate calculations are not the effective couplings of the axions to $W$- or $Z$-bosons, or to quarks, at least not in a proper technical sense.

The organization of this paper is as follows. In the next section we briefly review
the standard axion scenarios and the role of the Peccei-Quinn $U(1)_{PQ}$ symmetry
in them, also sketching why and how our proposal differs from the standard scenario.
In section~3 we summarize the basic features of the CSM needed for our calculation;
an alternative Lagrangian (`picture') giving the same results and  based on a redefinition of the
fermionic fields by a global $U(1)_{B\!-\!L}$  rotation is presented in section~4.
The requisite `neutrino technology', based on the systematic use of Weyl spinors and
off-diagonal neutrino propagators, is developed in the following section, followed by a  brief
discussion of neutrino mass matrices in section~6; here we also mention some
interesting physical implications, such as the possibility of light neutrino decays via
emission of `soft axions' (or `soft Majorons'). Section~7 deals with the issue of
$aZ$ mixing and the role of anomaly cancellations in the SM in the present proposal.
The remaining sections address the main topic of this paper, with the
calculation of the neutrino triangle diagram in section~8, the axion-quark diagrams in
section~9, and finally the axion-gluon vertex in section~10.

A main technical novelty is our consistent use of two-component (Weyl) spinors
for loop computations with neutrinos on the internal lines. Indeed, our results would
be rather cumbersome to rephrase in terms of 4-component (Dirac) spinors for the
neutrinos, so this formalism affords an efficient and elegant method for
computing higher order corrections in the presence of significant mixing between
the neutrino components. However, for those parts of the diagrams {\em not} involving
neutrinos, there is not much of a technical simplification, so we will return to the use
of 4-spinors for the quark loops in the last section. A short appendix explaining our
notations and conventions for two-spinors is included at the end (see also \cite{BW}).

\section{Axions: the `standard model', and beyond}
To contrast our proposal with current ideas and to make our
presentation self-contained, let us first recall some key features of
axion models leading to the standard Lagrangian governing the coupling
of axions to gauge fields
\beq\label{Lagg}
\cL_{\rm axion} (x) = \frac1{f_a} \frac{\al_s}{4\pi} \,
a(x)\, {\rm Tr} \, G^{\mu\nu} \tilde{G}_{\mu\nu}(x)
\eeq
(see also \cite{DiV} for a general introduction, and
\cite{Sikivie} for a review of the phenomenology of axions).
As commonly argued \cite{PQ,WW}, such a coupling can be generated if $a(x)$ is
a Goldstone boson associated with a spontaneously broken global $U(1)$ symmetry.
As we briefly explain below, this symmetry must furthermore be assumed {\em anomalous}
in order to give rise to the above term. This, then, is the famous Peccei-Quinn
symmetry $U(1)_{PQ}$, which is necessarily an {\em extra} global symmetry
beyond the known global symmetries already present in the SM. The strong
CP problem is then solved by arguing that the vacuum expectation value
of $a(x)$ dynamically adjusts itself to the value $\langle a \rangle =0$ \cite{VW}.

There are several axion models in the literature, all of which require extra
and so far unobserved new particles and scales in order to render the axion
`invisible' \cite{Kim,Shif,Dine,Zhit}; see also \cite{Khlopov1,Khlopov2,Lindner1}
where the breaking of the Peccei-Quinn symmetry is linked to neutrino physics.
Here, we briefly describe only the simplest
model of this type. This consists in enlarging the SM by a set of heavy
color triplet quarks $Q^a \equiv (Q^a_{L \al}, \bar{Q}_R^{\dal a})$
(where $a,...$ are $SU(3)_c$ indices), which couple in the standard way to
QCD, but do not transform under electroweak $SU(2)_w\times U(1)$, and
hence carry no electromagnetic charge. In addition we introduce a
{\em complex} scalar field $\phi$ which is a color and electroweak singlet, and
couples to the new quarks via (in terms of two-component Weyl-spinors)
\beq\label{axgg}
\cL = \phi \,\eps^{\al\be} Q^a_{L\al} Q_{R \be a} + \hc
\eeq
with the convention that a lower color index $a$ transforms in the $\bar{\bf{3}}$
of $SU(3)_c$. At the classical level, the full Lagrangian (including the terms we
have not written) is invariant under the global $U(1)_{PQ}$ symmetry
\beq
\phi \rightarrow e^{2\ri\om}\phi \; , \quad
Q^a_{L\al} \rightarrow e^{- \ri\om} Q^a_{L\al}\;,\quad
Q_{R \al a} \rightarrow  e^{-\ri\om} Q_{R\al a}
\eeq
In 4-spinor notation this is equivalent to the transformation
$Q^a\rightarrow e^{-\ri\om\ga^5} Q^a$, whence the $U(1)_{PQ}$ symmetry is chiral.
If we now represent the scalar field as
\beq\label{phi}
\phi(x) = \varphi (x) \exp \big(\ri a(x)/\sqrt{2} \mu \big)
\eeq
and assume the $U(1)_{PQ}$ to be spontaneously broken with vacuum
expectation value $\vev$, we can make the new quarks as heavy as we like
by arranging $\vev$ to be large enough; this is, in fact, necessary
in order to avoid immediate conflict with observation. The phase factor $a(x)$,
on the other hand, is turned into a Goldstone boson. The coupling (\ref{Lagg})
then effectively arises via the diagram
\begin{center}
\includegraphics[width=13cm,viewport= 100 620 540 720,clip]{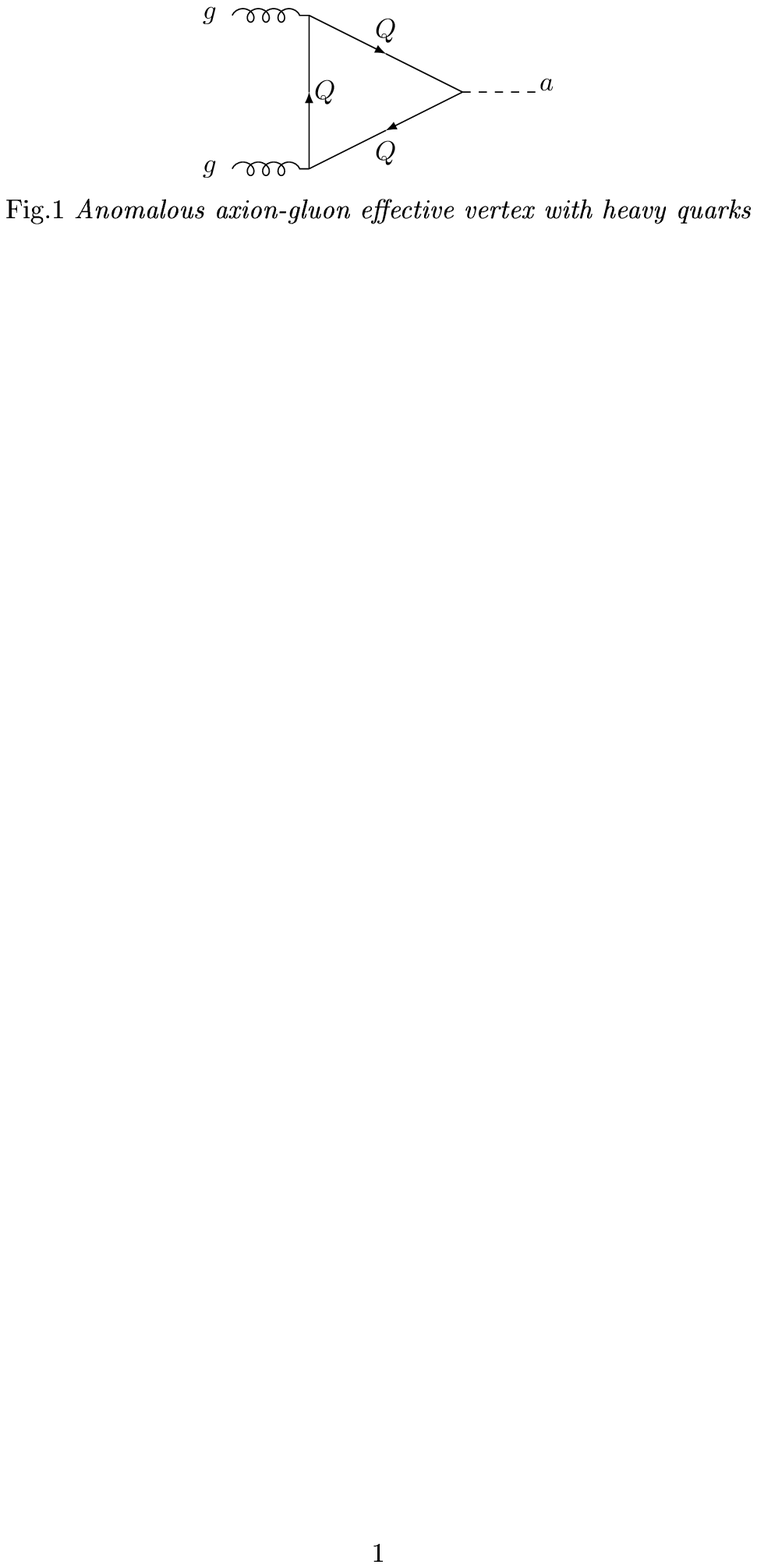}
\end{center}
Observe that this diagram gives a non-zero result precisely because
of the anomaly: if this anomaly were cancelled by further contributions
from other fields the effect would disappear. This is the reason why
in all available models the $U(1)_{PQ}$ is necessarily anomalous.

By and large, all existing axion scenarios are variants or more elaborate
extensions of this basic model. For instance, if one wants the axion to
also couple to photons one must allow the new scalar field to also
transform under electroweak symmetries, e.g. by introducing further
Higgs doublets into the theory, as happens to be the case for supersymmetric
extensions of the SM, thereby making room for extra global symmetries.
A feature common to all these scenarios is that one must tune the relevant scales
{\em by hand} to very large (or small) values in order to achieve the required
invisibility of the axion.

As a general feature we note that (\ref{Lagg}) is the only possibility to reconcile
unbroken $SU(3)_c$ gauge invariance with the fact that $a(x)$ is a Goldstone boson,
which requires all its couplings to be via derivatives. This is because the
gluonic topological density can be represented as the divergence of a
{\em local} quantity, the Chern-Simons current $\cJ^\mu_{CS}$, according to
\beq\label{CS}
{\rm Tr} \, G^{\mu\nu} \tilde{G}_{\mu\nu} = \partial_\mu \cJ^\mu_{CS}
\eeq
whence the coupling (\ref{Lagg}) is equivalent by partial integration to the
derivative coupling $\partial_\mu a \,\cJ^\mu_{CS}$. This argument and the
Goldstone nature of $a(x)$ imply in particular that, with unbroken gauge invariance, couplings $\propto a\, {\rm Tr}\, G^{\mu\nu} G_{\mu\nu}$ (or higher dimension couplings of this kind) can never be generated. With regard to the unbroken $SU(3)_c \times U(1)_{em}$ gauge sector,
that is, its coupling to gluons and photons, the axion $a(x)$ therefore effectively
behaves like a pseudoscalar.

In this paper we present a scenario which differs from the one above in some essential
regards. {\em Nevertheless the basic mechanism of identifying the axion $a(x)$ with
the Goldstone boson of a  spontaneously broken global $U(1)$ symmetry and of generating
the desired coupling (\ref{Lagg}) via a loop diagram
remains in force}. Yet, we here do {\em not} assume $\phi(x)$ to be a new
scalar field; rather, we propose to identify it with the extra field already present in the CSM,
whose non-vanishing expectation value provides the Majorana mass term for the
right-chiral neutrinos, while its phase is identified with the so-called Majoron, {\it alias} the Goldstone boson of spontaneously broken lepton number symmetry. This field is
a natural ingredient of any `minimal' extension of the SM where this mass term is not
simply put in by hand via an explicit Majorana mass term, and generally  occurs
in classically conformal versions of the SM with right-chiral neutrinos. Consequently,
we claim that the role of the Peccei-Quinn symmetry can be taken over by a known
symmetry, lepton number symmetry $U(1)_L$, whose associated Goldstone boson
thus becomes identified with the axion. 

\begin{center}
\includegraphics[width=13cm,viewport= 100 620 540 720,clip]{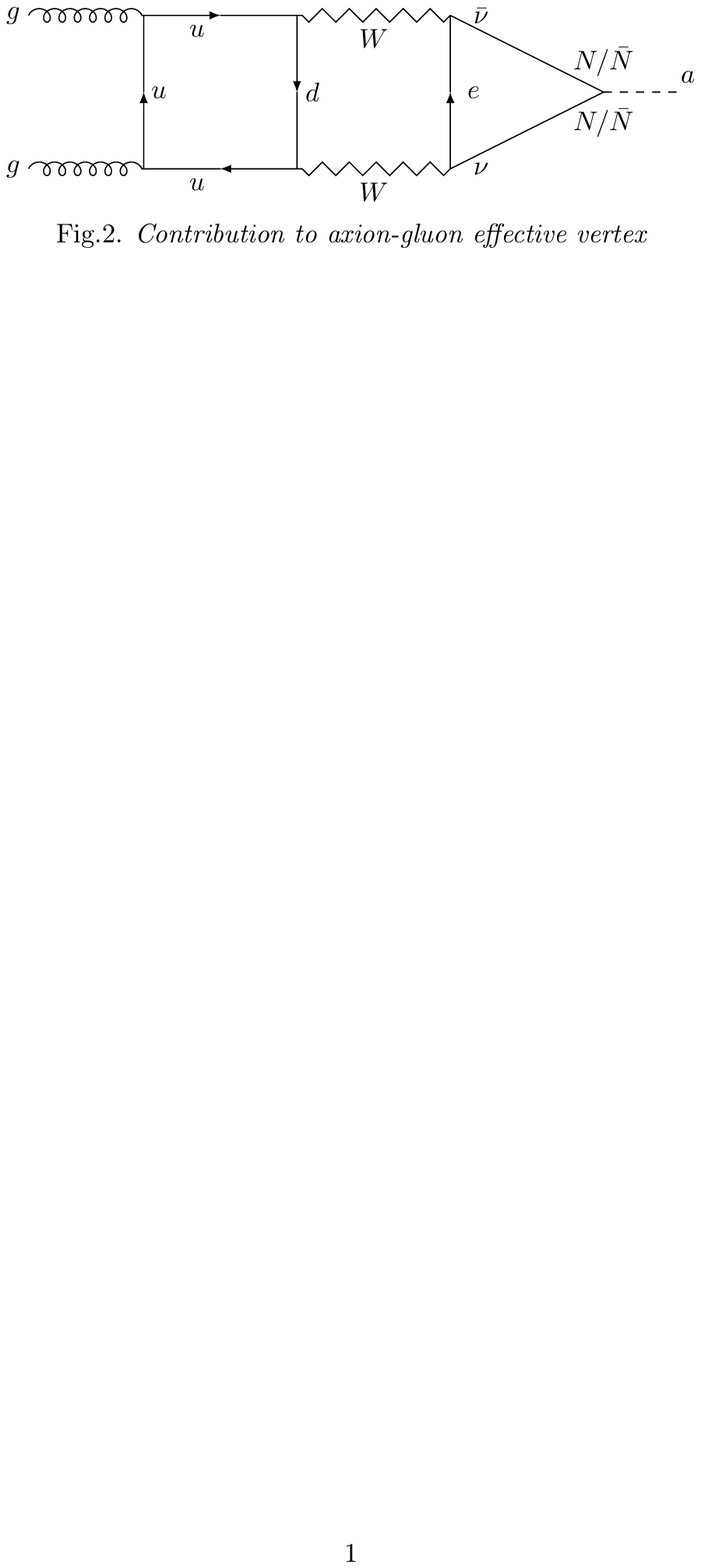}
\end{center}

Evidently, our proposal runs counter to accepted wisdom as outlined above in several
aspects. First of all, we here propose to cancel the strong CP term not by a strictly local
term like (\ref{Lagg}), but by a {\em non-local} term which arises as a contribution
to the effective action via the diagrams in Fig.~1, and which reduces to the local expression
(\ref{Lagg}) only in the long distance (infrared) limit. In fact, (\ref{Lagg})
is the {\em only} term that on general grounds can survive this limit:  like for the standard
axion scenarios, the Goldstone nature of $a(x)$ and the unbroken $SU(3)_c$ gauge
invariance imply that (\ref{Lagg}) is the only possible {\em local} coupling that can
emerge. These two principles are, however, not at all in contradiction with
{\em non-local} interactions, which an expansion in momenta of the full effective can and
does produce.

Secondly, lepton number symmetry $U(1)_L$ is non-chiral, and $(B\!-\!L)$ symmetry
is non-anomalous. As we will explain in much detail below, the effective coupling
(\ref{Lagg}) nevertheless does arise in the long wave length limit due to the very
special features of the neutrinos and their mixing. More specifically, the anomalous
triangle involving heavy quarks (cf. Fig.1) is here replaced by a set of
three-loop diagrams (an exemplary one of which is displayed in Fig.2) with a
neutrino triangle diagram at one end. As we will show, the latter, though UV finite, can
give rise in the long distance limit to anomaly-like amplitudes due to the mixing of neutrino components and the fact that parity is maximally violated in the SM, and therefore plays the
same role as the anomalous triangle in the usual models.
One difference should be noted, though: while we make do here
with an uncorrected quark box on the left side of the diagram, the actual value of
the effective coupling could is affected by higher order QCD corrections,
because $\al_S$ is large at small momenta (there is no analog of the
Adler-Bardeen Theorem in the present situation). Although such corrections could change
our prediction for $f_a$, this is not a main concern here because the precise value
of $f_a$ does not matter so much for the  solution of the strong CP problem, as long
as it is different from zero, as shown to be the case by our explicit calculation.

An altogether different option starting from the present model would be to promote
$(B\!-\!L)$ to a {\em local} symmetry by gauging $U(1)_{B\!-\!L}$ (see \cite{BPY} and
references therein, and \cite{Ito} for more recent work).
In this case,  the Majoron would be swallowed by  the $U(1)_{B\!-\!L}$
gauge boson to give it a mass, and could thus not become an axion. The possibility
of gauging $U(1)_{B\!-\!L}$ is naturally suggested by the fact that the $U(1)_{B\!-\!L}$
symmetry is non-anomalous,  and this is apparently the point of view taken in much of the pertinent literature on this subject, where the possibility of keeping the Majoron as an `uneaten'
Goldstone boson was never seriously considered, and effective vertices of the type
analyzed in this paper were not computed (though known to exist).

\section{Basic properties of the CSM}

We refer readers to \cite{EPP,Pok} for the SM Lagrangian, enlarged by a family
triplet of right-chiral neutrinos, and to \cite{Ramond} for an up-to-date overview.
The scalar sector of the model consists of the SM Higgs doublet and in addition a complex electroweak singlet scalar field $\phi(x)$ coupled to the right-chiral neutrinos via the usual Majorana term. Consequently the Yukawa couplings are given by
\bea\label{L}
-\cL_{\rm Y}\!\!\!&=&\!\!\!
\overline{L}^i\Phi Y_{ij}^E E^j  +
\overline{Q}^i\Phi Y_{ij}^D D^j+\overline{Q}^i\eps\Phi^* Y_{ij}^U U^j\nn\\
&& + \, \overline{L}^i\eps\Phi^\ast Y_{ij}^\nu N^j+\frac12 \phi N^{i T}
\cC Y^M_{ij} N^j+{\rm h.c.}
\eea
Here $Q^i$ and $L^i$ are the left-chiral quark and lepton doublets, {\it viz.}
\beq\label{Qi}
Q^i \equiv \left(\begin{array}{c} u^i_L\\[1mm]
                                                        d^i_L \end{array}\right)\;,\quad
 L^i \equiv \left(\begin{array}{c} \nu^i_L\\[1mm]
                                                        e^i_L \end{array}\right)
\eeq
while $U^i$ and $D^i$ are the right-chiral up- and down-like quarks, $E^i$
the right-chiral electron-like leptons, and $N^i\equiv\nu^i_R$ the right-chiral
neutrinos (we suppress all indices except the family indices $i,j=1,2,3$; summation
over double indices will be understood\footnote{For clarity, we will, however, suspend
  the summation convention for flavor indices occasionally, and also use capital letters
  $I,J,\dots$ for the quark flavors in sections 8ff.}). The Yukawa matrices $Y^\sharp_{ij}$
in (\ref{L}) are general complex 3-by-3 matrices, except for $Y^M_{ij}$ which is complex
and symmetric. $\Phi$ is the usual Higgs doublet, and $\phi$ is the new complex scalar field;
note that this field has no direct couplings to the other SM fields (but such couplings will arise either through mixing or higher loop effects).\footnote{More generally, one might replace
  the `flavor neutral' complex scalar $\phi(x)$ in (\ref{L}) by a field $\phi_{ij}(x)$ transforming
  as a sextet under a horizontal (family) symmetry $SU(3)_H$ \cite{Khlopov1}.}

Since we are going to use two-spinors mostly let us spell out the relation with
the four-dimensional spinors, in order to be completely explicit about our notation
(see also the Appendix):
\beq\label{ui}
u^i \equiv \left(\begin{array}{c} u^i_{L\al}\\[1mm]
                                      \bu_R^{i\dal} \end{array}\right)\;,\;
d'^i \equiv\left(\begin{array}{c} d'^i_{L\al}\\[1mm]
                                       \bd'^{i\dal}_R \end{array}\right)\;,\;
e^i \equiv \left(\begin{array}{c} e^i_{L\al}\\[1mm]
                                       \bel_R^{i\dal} \end{array}\right)\;,\;
\nu^i \equiv \left(\begin{array}{c} \nu^i_{\al}\\[1mm]
                                         \bN^{i\dal} \end{array}\right)
\eeq
where we find it convenient to define the down-quarks with an extra rotation
involving the CKM matrix $V^{ij}$
\beq\label{didef}
d'^i_{L\al} \equiv V^{\dagger ij} d^j_{L\al},\ \ \bd'^{i\dal}_R\equiv V^{ij}\bd_R^{j\dal}
\eeq
see (\ref{YD}) below. Also for later convenience, we have
chosen to label the independent neutrino Weyl spinors by {\em different} letters
$\nu$ and $N$, while for the other fermions we adopt the
convention of labeling the independent two-component Weyl spinors by the subscripts
$L$ and $R$. Using the formulae of Appendix A, one then has the usual relations
between 4-spinor and two-spinor expressions, such as for instance
$\bu^i u^i \equiv u_L^{i\al} u^i_{R\al} + \bu^i_{R\dal} \bu^{i\dal}_L
=   u_R^{i\al} u^i_{L\al} + \bu^i_{L\dal} \bu^{i\dal}_R$ or
$\bu^i \ga^5 u^i \equiv u_L^{i\al} u^i_{R\al} - \, \bu^i_{R\dal} \bu^{i\dal}_L$,
and so on.

As is well known, one can use global unitary redefinitions of the fermion fields
$L^i, E^i, Q^i, U^i$ and $N^i$ to transform the Yukawa matrices $Y_{ij}^E$, $Y_{ij}^U$ and $Y_{ij}^M$ to real diagonal form. To simplify the remaining (still general complex) matrices $Y_{ij}^D$ and $Y^\nu_{ij}$ we still have three phase $[ = U(1)^3]$ rotations on $(L^i,E^i)$, and another three on $(Q^i,U^i)$ at our disposal, as well as the remaining unitary rotation of the down quarks $D^i$. The latter can be used to represent the matrix $Y^D$ as
\beq\label{YD}
Y^D = V R^D V^\dagger
\eeq
where $R^D$ is real diagonal and $V$ is the CKM matrix (with three angles and one
phase); the extra factor $V^\dagger$ here is related to our definition
(\ref{didef}) for the down quarks. Finally, exploiting all remaining phase
rotations, the neutrino Yukawa matrix can be brought to the form
\beq\label{Ynu}
Y^\nu = K_\nu^R P^R_\nu R_\nu K_\nu^L P_\nu^L
\eeq
where $R_\nu$ is real diagonal, $K_\nu^{L,R}$ are CKM-like matrices (each with three angles
and one phases), and $P^{L/R}_\nu$ are diagonal phase matrices of unit determinant.
The matrix $Y^\nu$ thus represents altogether 15 free parameters (three
real parameters, six angles and six phases). Moreover, and in contrast to
the CKM matrix, $K_\nu^{L,R}$ may (and $K_\nu^L$ is expected to) exhibit strong mixing.

The scalar sector is governed by the usual Lagrangian
\beq\label{PHphi}
\cL_{\rm scalar} =  - D^\mu \Phi^\dagger D_\mu \Phi
 - \partial^\mu \phi^* \partial_\mu\phi - \cP (H, \vp)
 \eeq
where $H^2\equiv \Phi^\dagger\Phi$ and $\vp^2 \equiv \phi^* \phi$. We do not further specify the potential here, but only make the usual assumption that the fields $\Phi$ and $\phi$ acquire non-vanishing vacuum expectation values by spontaneous symmetry breaking (which may occur either via the standard Mexican hat potential, or via a Coleman-Weinberg type breaking from a classically conformal Lagrangian; in the latter case, $\cP(H,\vp)$ would also contain logarithmic terms). The breaking in particular entails Dirac- and Majorana mass matrices for the neutrinos
\beq\label{mM}
m_{ij} := \langle H\rangle \, Y^\nu_{ij} \;\; , \quad M_{ij} := \vev \, Y^M_{ij}
\eeq
where $H^2 \equiv \Phi^\dagger \Phi$. With our choice of phases $M_{ij}$
becomes a real diagonal 3-by-3 matrix. The matrix $m_{ij}$ stays complex,
but from (\ref{YD}) we see that it can be represented in the form $m=Vm^DV^\dagger$,
with the diagonal matrix $m^D = \langle H\rangle R^D$.
As we already pointed out the vacuum expectation value $\vev$ need not be several
orders of magnitude above the electroweak scale, but can be taken of the same order
as $\langle H\rangle$; the smallness of the masses of light neutrinos is then achieved
by taking the neutrino Yukawa couplings to be $Y^\nu \sim \cO(10^{-5})$.~\footnote{In
  this case, the heavy neutrinos generally have masses $< \cO(1\,{\rm TeV})$.
  Also, there are then no large corrections to the Higgs mass from the exchange of
heavy neutrinos \cite{MSV}.}
We also note that with the representation (\ref{phi}) and $\vev\neq 0$
the canonical normalization of the kinetic term is obtained with $\mu = \vev$
and $\vp = \vev + (1/\sqrt{2}) \vp'$ such that
\beq\label{phikin}
- \partial^\mu \phi^* \partial_\mu\phi = - \frac12 \partial^\mu\vp' \partial_\mu\vp'
    - \frac12 \partial^\mu a \,\partial_\mu a + \cdots
\eeq
The field $\vp'(x)$ will not play a significant role in the remainder.

Besides the (local) $SU(3)_c\times SU(2)_w \times U(1)_Y$ symmetries,
the CSM Lagrangian admits two {\em global} $U(1)$ symmetries, lepton number
symmetry $U(1)_L$ and baryon number symmetry $U(1)_B$. These are, respectively,
associated with the Noether currents
\bea\label{JL}
\cJ^\mu_L &:=&  \overline{L}^i\ga^\mu L^i
+\overline{E}^i\ga^\mu E^i+\overline{N}^i\ga^\mu N^i
-2 \ri \phi^\dagger\! \stackrel{\leftrightarrow}{\partial^\mu} \!\phi      \nn\\
&\equiv&  \bar{e}^i \ga^\mu e^i + \bar\nu^i \ga^\mu \nu^i
-2 \ri \phi^\dagger\! \stackrel{\leftrightarrow}{\partial^\mu} \!\phi
\;\equiv \;\JJ_L^\mu -2 \ri \phi^\dagger\! \stackrel{\leftrightarrow}{\partial^\mu} \!\phi
\eea
and
\bea\label{JB}
\cJ^\mu_B &:=&
\frac13\,\overline{Q}^i\ga^\mu Q^i
+\frac13\,\overline{U}^i\ga^\mu U^i+\frac13\,\overline{D}^i\ga^\mu D^i\nn\\
&\equiv& \frac13 \bu^i\ga^\mu u^i + \frac13 \bd^i\ga^\mu d^i
  =  \frac13 \bu^i\ga^\mu u^i + \frac13 \bd'^i\ga^\mu d'^i
\eea
(where by $u^i,d^i,e^i$ and $\nu^i$ we here denote the full Dirac 4-spinors, see above).
Evidently both currents are purely vector-like. Furthermore, the scalar $\phi$
carries two units of lepton number charge, hence lepton charge can `leak' from the
fermions into the scalar channel. For spontaneously broken lepton number
the {\em total} current $\cJ_L^\mu$ remains conserved, but its fermionic part
is not conserved (even at the classical level) because
\beq\label{JL1}
\partial_\mu \JJ_L^\mu =
     -  \ri M_{ij} \big( N^{i\al} N^j_\al - \bar{N}^i_{\dot\al} \bar{N}^{j\dot\al}\big) \neq 0
\eeq
where the mass matrix $M_{ij}$ defined in (\ref{mM}) is non-vanishing for $\vev\neq 0$.
An important (and well known) fact is that the $(B\!-\!L)$ current is free of anomalies,
whereas $\cJ_L^\mu$ and $\cJ_B^\mu$ are anomalous separately.

Finally, we write out those terms in the CSM Lagrangian relevant for the computation.
After symmetry breaking and diagonalization of all mass matrices except neutrino ones, the electron and quark mass terms read
\beq
\cL_{\rm mass} =
- \sum_i\Big(m_{e_i} \bar{e}^i e^i + m_{d_i} \bar{d}'^i d'^i+ m_{u_i} \bar{u}^i u^i\Big)
\eeq
where, of course, $\{ m_{e_i}\}\equiv (m_e, m_\mu, m_\tau)$, and so on.
Using two-spinors the mass terms for neutrinos and the axion interactions read
\bea\label{Lbreak}
\cL_{\rm int}^{(1)} &=&  - \Big(\nu^{i\al} m^{ij} N^j_\al + \bar\nu^i_{\dot\al} (m^*)^{ij}
\bar N^{j\dot\al} +  \frac12 N^{i\al}M^{ij}N^j_\al + \frac12\bar{N}^i_{\dot\al}
(M^*)^{ij}\bar{N}^{j\dot\al} \Big) \nn\\
&& \qquad - \, \frac{\ri a}{2\sqrt{2}\vev} \Big(N^{i\al} M^{ij} N^j_\al - \bar N^i_{\dot\al} (M^*)^{ij} \bar N^{j\dot\al}\Big)
\eea
The interactions of the leptons with $W$ and $Z$ bosons are given by
\bea
\cL_{\rm int}^{(2)} &=&
 - \, \frac{g_2}{\sqrt{2}}\, W_\mu^+ \bnu^i_\dal \bar\si^{\mu\dal\be} e^i_{L\be}
 - \frac{g_2}{\sqrt{2}}\, W_\mu^- \bar{e}^i_{L\dal} \bar\si^{\mu\dal\be} \nu^i_{\be} \\
 &&  \!\!\!\!\! \!\!\!\!\! \!\!\!\!\! \!\!\!\!\! \!\!\!\!\!
 - \, \frac{g_2}{\cos\theta_w} Z_\mu
   \left[\frac12 \left( \bnu^i_\dal \bar\si^{\mu\dal\be} \nu^i_\be -
          \bar{e}^i_{L\dal} \bar\si^{\mu\dal\be} e^i_{L\be}\right)
 + \sin^2 \theta_w \left( \bar{e}^i_{L\dal} \bar\si^{\mu\dal\be} e^i_{L\be} +
      e^{i\al}_R \si^\mu_{\al\dbe} \bar{e}^{i\dbe}_R
          \right)\right]     \nn
\eea
while for the quarks they read
\bea
\cL_{\rm int}^{(3)} &=&
 - \, \frac{g_2}{\sqrt{2}}\, W^+_\mu V^{ij} \bar{u}^i_{L\dal} \bar\si^{\mu\dal\be} d'^j_{L\be} -
\frac{g_2}{\sqrt{2}}\, W^-_\mu (V^\dagger)^{ij} \bar{d}'^i_{L\dal} \bar\si^{\mu\dal\be} u^i_{L\be}    \\
&&  \!\!\!\!\!\!\!\!\!\!\!\!\!\! \!\!\!\!\!\!\!\!\!\!
 - \, \frac{g_2}{\cos\theta_w} Z_\mu
   \left[\frac12 \left( \bar{u}^i_{L\dal} \bar\si^{\mu\dal\be} u^i_{L\be} -
          \bar{d}'^i_{L\dal} \bar\si^{\mu\dal\be} d'^i_{L\be}\right) \right. \nn \\
&& \left. \!\!\!\!\!\!\!\!\!\!+ \sin^2 \theta_w \left(\frac13 \bar{d}'^i_{L\dal} \bar\si^{\mu\dal\be} d'^i_{L\be} + \frac13
     d'^{i\al}_R \si^\mu_{\al\dbe} \bar{d}'^{i\dbe}_R - \frac23 \bar{u}^i_{L\dal} \bar\si^{\mu\dal\be} u^i_{L\be} - \frac23
     u^{i\al}_R \si^\mu_{\al\dbe} \bar{u}^{i\dbe}_R
          \right)\right]     \nn
\eea
with the CKM matrix $V^{ij}$. Here $g_2$ is, of course, the $SU(2)_w$ gauge coupling.

\section{Alternative Picture}

Although we will use (\ref{L}) to calculate the effective couplings, we should
mention that there exists an equivalent approach (or `picture')
that emphasizes the fact that the phase of the complex field $\phi(x)$
becomes a Goldstone boson after spontaneous symmetry breaking.
Using (\ref{phi}) and redefining the fermionic fields according to
\bea\label{phidef}
\big(L^i(x),E^i(x),N^i(x)\big)\; &\to& \;
\exp\left(-\frac{\ri a(x)}{2\sqrt{2}\mu}\right)\big(L^i(x),E^i(x),N^i(x)\big)\; , \nn\\
\big(Q^i(x),U^i(x),D^i(x)\big)\; &\to& \;
\exp\left(\frac{\ri a(x)}{6\sqrt{2}\mu}\right)\big(Q^i(x),U^i(x),D^i(x)\big)
\eea
we can replace the complex field $\phi(x)$ by the {\em real}  field $\varphi(x)$.
The phase $a(x)$ then occurs only via its derivatives in the re-defined Lagrangian,
as appropriate for a Goldstone boson. Defining the total $(B\!-\!L)$ current
\beq
\cJBL^\mu :=\cJ^\mu_B - \cJ^\mu_L \,\equiv \, \JJBL^\mu
 \, + 2 \,   \ri \phi^\dagger\! \stackrel{\leftrightarrow}{\partial^\mu} \!\phi
\eeq
and using (\ref{phi}) with $\mu = \vev \neq 0$, this current assumes the universal form
\beq
\cJBL^\mu = \JJBL^\mu - \frac{\vev}{\sqrt{2}} \, \partial^\mu a
\eeq	
for a current with a Goldstone boson $a(x)$, with corresponding Lagrangian
\beq\label{Goldstone}
\cL_{\rm Goldstone} \, = - \frac12 \partial_\mu a\partial^\mu a
      + \, \frac{\sqrt{2}}{\vev} \partial_\mu a \, \JJBL^\mu
\eeq
Varying this Lagrangian w.r.t. $a(x)$, the resulting equation of motion
implies the conservation of the total current in the form
\beq\label{conservation}
 \frac{\vev}{\sqrt{2}} \, \Box a \, - \,  \partial_\mu \JJBL^\mu = 0
\eeq
It is worth emphasizing that the redefinition (\ref{phidef})
is also {\em well-defined quantum mechanically},  precisely because the
$(B\!-\!L)$ current is anomaly free in the SM, unlike the currents
$\cJ^\mu_B$ and $\cJ^\mu_L$ separately --- this was our reason for including
the quark fields in the redefinition (\ref{phidef}). Therefore the change of
variables (\ref{phidef}) does not affect the fermionic functional measure, ensuring
the mutual consistency of the two pictures also at the quantum level. In other words, it
does not make any difference whether we base our calculation on the Yukawa
Lagrangian (\ref{L}) or on the vertex $\partial_\mu a\, \JJBL^\mu$: the final result
must be the same.

Importantly, the conservation condition (\ref{conservation}) for the generalized
current says nothing about {\em how}  the two contributions $\partial_\mu \JJBL^\mu$
and $\Box a$ conspire to produce overall current conservation by using the classical or
quantum equations of motion. All it says is that, whenever $\Box a \neq 0$,
there must be a corresponding contribution  to $\partial_\mu \JJBL^\mu\neq 0$
for (\ref{conservation}) to be satisfied. Thus
\beq\label{ME0}
\Box a = \frac{\sqrt{2}}{\vev}\,  \rX \;\;\Rightarrow \quad
\partial_\mu \JJBL^\mu =  \rX
\eeq
for the full classical or quantum equations of motion. At the classical level
and with spontaneous symmetry breaking we have
\beq\label{JJBL}
\partial_\mu \JJBL^\mu = \rX
     =  \ri M_{ij} \big( N^{i\al} N^j_\al - \bar{N}^i_{\dot\al} \bar{N}^{j\dot\al}\big)
\eeq
as a consequence of (\ref{JL1}). One easily checks that the Lagrangian (\ref{L})
gives rise to a corresponding contribution to $\Box a$ which just cancels the
above term in the divergence of the total current to give $\partial_\mu \cJBL^\mu =0$.

This cancellation mechanism persist at the quantum level. Here, we determine the
effective couplings  of $a(x)$ to other fields by evaluating the matrix elements
\beq
\big\langle \Psi \big| a \, \partial_\mu\JJBL^\mu \big| a\big\rangle_{\rm 1PI}
\eeq
where $|\Psi\rangle$ can be any (multi-particle) state involving excitations other
than $a$, and where the subscript indicates that we amputate the external legs
in the usual fashion. To get the corresponding contribution to the quantum equation
of motion we stick on the classical fields $\chi_1, \dots, \chi_n$ associated to the
particular state $|\Psi\rangle$. Schematically, this turns the effective
equation of motion for $a$ (resulting from the full quantum effective action) into
\beq
\frac{\vev}{\sqrt{2}}\,\Box a \,  +
\cdots + \, \chi_1\cdots\chi_n \big\langle\Psi\big|\partial_\mu \JJBL^\mu \big|0\big\rangle_{\rm 1PI}
   + \cdots = 0
\eeq
where the new terms above and beyond (\ref{JJBL}) appear at higher orders in
$\hbar$ and represent the quantum corrections to the classical equation of motion.
Note that the terms involving $\chi$'s are in {\em non-local} in general, but we are here
primarily interested in the {\em quasi-local} approximation where we integrate out all massive fields, and look only at long wave-length (low momentum) excitations. Furthermore, in the
case at hand, all these corrections are due to neutrino mixing, that is, the non-vanishing
r.h.s. of (\ref{JJBL}); if it did vanish, $a(x)$ would be a free field.

In this paper, we will compute various such quantum corrections to the classical
equation of motion $\Box a = \cdots$ using the Lagrangian (\ref{L}), but the above
considerations show that the `picture' of this section would give the same results.
Schematically, the corrections (ordered in powers of $\hbar$) are of the form
\beq
\hrX = \rX + \hbar (WW + ZZ) + \hbar^2 \big( \bar{f}f + F\tF)
 + \hbar^3 G\tG + \cdots
\eeq
where the letter $f$ stands for quarks and electrons, and by $F,W,Z$ and $G$,
we schematically denote the field strengths of the associated vector bosons.
{\em There is nothing that forbids such local couplings to appear in the long distance
limit.} Equally important, the different terms `kick in' at different energies, in accord
with what we said in the introduction to this paper.  For instance,
at low momenta only the $\hbar^2 F\tF$ and $\hbar^3 G\tG$ terms are present.
At higher energies (integrating out fewer degrees of freedom), these vertices
`dissolve' to become non-local, while the $W\tilde{W}$ contribution is still
effectively local at energies $\sim M_W$. For the $aWW$ vertex, the equivalence of
the two pictures has now been explicitly confirmed in \cite{LMN2}.

\section{Neutrino propagators}

Using two-component spinors, and after symmetry breaking, the free part
 of the neutrino Lagrangian is
\bea
 \cL &=& \frac{\ri}{2}\left(\nu^{\al i} \dslash_{\al\dot\be}\bar\nu^{\dot\be i} +
   \bar\nu^i_{\dot\al}\dslashbar^{\dot\al\be}\nu^i_\be +
 N^{\al i} \dslash_{\al\dot\be}\bar N^{\dot\be i} +
 \bar{N}^i_{\dot\al}\dslashbar^{\dot\al\be}N^i_\be\right) \nn \\
&& \quad
- \, m_{ij}\nu^{\al i}N_{\al}^j - m^*_{ij} \bar\nu_{\dot\al}^i \bar{N}^{\dot\al j}
-\frac12 M_{ij}  N^{\al i}N_\al^j  -\frac12  M^*_{ij} \bar{N}_{\dot\al}^i \bar{N}^{\dot\al j}
\label{kinterms}
\eea
where, as before, the indices $i,j=1,2,3$ label the family (the sum over which is
understood). As already pointed out above, by making a unitary rotation on the fields
$N^i$, we can bring the Majorana mass matrix $M_{ij}$ to real diagonal form,
with strictly positive eigenvalues, {\it viz.}
\beq\label{Mdiag}
M_{ij} = \delta_{ij} M_j \;, \quad M_j > 0
\eeq
By contrast, the Dirac mass matrix $m_{ij}$ remains a general complex 3-by-3 matrix.
Let us emphasize, however, that the results to be presented do not depend on
choices of phases or specific representations of these matrices, which are
therefore mainly a matter of convenience.

At this point we have two options. The first is to diagonalize the neutrino mass
term (\ref{kinterms}) by re-defining the neutrino spinors
\beq\label{MNdiag}
\nu_\al'^i = \cU_1^{ij}\nu^j_\al  + \cU_2^{ij} N^j_\al \;, \quad
N'^i_\al= \cU_3^{ij}\nu^j_\al  + \cU_4^{ij} N^j_\al
\eeq
by means of a unitary 6-by-6 matrix
\beq\label{U}
\cU = \begin{bmatrix}
             \cU_1 & \cU_2 \\[1mm]
             \cU_3 & \cU_4
\end{bmatrix} \; , \qquad \cU \cU^\dagger = {\bf{1}}
\eeq
preserving the kinetic terms in (\ref{kinterms}). In this way the quadratic terms become
diagonal, but the vertices coupling neutrinos to SM fields become off-diagonal. We will refer
to this description as the `propagator picture'. In the following section, we will discuss the
neutrino mass matrices in somewhat more detail, and also return to eqns.~(\ref{MNdiag}) and (\ref{U}).

Alternatively, however, one may adopt another description (the `vertex picture')
where the vertices remain diagonal (i.e. in the form of the
Lagrangian (\ref{Lbreak})), but the propagators are off-diagonal. This `vertex picture' has the advantage that we can simply use the original SM Lagrangian for the interaction vertices. For this purpose we need to invert the (manifestly hermitian)
operator
\beq\label{KinOp}
\cK =
\begin{bmatrix}
\ri\dslashbar^{\dot\al\be} & 0 & 0 & -m^* \delta^{\dot\al}_{\dot\be} \\[1mm]
0 & \ri\dslash_{\al\dot\be} & -m\delta_{\al}^{\be} & 0 \\[1mm]
0 & -m^\dagger \delta^{\dot\al}_{\dot\be} & \ri\dslashbar^{\dot\al\be} &
       -M\delta^{\dot\al}_{\dot\be}\\[1mm]
-m^T \de_{\al}^{\be} & 0 & -M \delta_{\al}^{\be} & \ri\dslash_{\al\dot\be}
\end{bmatrix}
\eeq
This operator is to be sandwiched between the multi-spinors
$\left[\bar\nu_{\dal},\nu^\al,{\bar N}_{\dal},N^\al\right]$ on the left and
$\left[\nu_{\be},\bar\nu^{\dot\be},N_\be,{\bar N}^{\dot\be}\right]^T$ on the right.
Because each entry here is a 3-by-3 matrix in family space, the operator $\cK$
is thus represented as a 12-by-12 matrix operator. It is convenient at this point to adopt the form (\ref{Mdiag}), and this will be assumed from now on, while the matrix $m$ is left in the general complex form. The matrix inversion can be performed in
momentum space by iterated use of the formula
\beq
\begin{bmatrix}
             A & B\\[1mm]
             C & D
\end{bmatrix}^{-1}   =
\begin{bmatrix}
A^{-1}(A+B(D - CA^{-1}B)^{-1}C)A^{-1}  & - A^{-1}B(D-CA^{-1}B)^{-1}\\[1mm]
- (D - CA^{-1}B)^{-1} CA^{-1} & (D - CA^{-1}B)^{-1}
\end{bmatrix}
\eeq
where the sub-matrices $A$ and $D-CA^{-1}B$ are assumed to be invertible (we can
arrive at  different forms of this identity by acting with the matrix $\begin{bmatrix}
             0& 1\\
             1 & 0
\end{bmatrix}$ on one or both sides of the original matrix and identifying sub-matrices in a different way).
Identifying
\beq
 A \equiv \begin{bmatrix}
             \PSB & 0\\
             0 & \PS
            \end{bmatrix} \;,\;
D \equiv \begin{bmatrix}
             \PSB & -M\\
             -M & \PS
            \end{bmatrix}\;,\;
B \equiv \begin{bmatrix}
             0 & -m^*\\
             -m & 0
            \end{bmatrix} \;,\;
C \equiv B^\dagger
            \;,
\eeq
we get the result for the inverse sub-matrix

\beq
\left(D-CA^{-1} B\right)^{-1} =
\begin{bmatrix}
      \;\cD(p)(p^2-m^T m^*) M^{-1} \pslash & \cD(p)p^2\\[2mm]
      \cD(p)^* p^2 & M^{-1}(p^2-m^\dagger m)\cD(p)\pslashbar\;
            \end{bmatrix}
\eeq
where the 3-by-3 matrix $\cD(p)$ is defined by
\beq\label{defDp}
 \cD(p) := \Big[ \big( p^2 - m^Tm^* \big)M^{-1}
    \big( p^2 - m^\dagger m \big) - M p^2 \Big]^{-1}\;=\;  \cD(p)^T
\eeq
A useful alternative form is
\bea
&& \!\!\!\!\!\!\!\!\!\!\!\!
\cD(p) := M^{1/2}\Big[p^4 \, -\,  p^2\big(M^2 + M^{1/2} m^T m^* M^{-1/2} +
    M^{-1/2} m^\dagger m M^{1/2}\big)   \nn\\
&& \qquad\qquad\qquad
     + \; M^{1/2} m^T m^* M^{-1} m^\dagger m M^{1/2}\Big]^{-1} M^{1/2}
\eea
Therefore (not forgetting an extra factor of $\ri$ in front) the results for the
propagator components read:
\bea\label{NProp}
\langle \nu^i_\al (x) \nu^{j\be} (y) \rangle &=& \ri \int\frac{d^4p}{(2\pi)^4}\left[m^* \cD(p)^* m^\dag\right]^{ij} \de_\al^\be \re^{-\ri p\, (x-y)} \\
\langle \nu^i_\al (x) \bar\nu^j_\dbe (y) \rangle &=& \ri\int\frac{d^4p}{(2\pi)^4}\left[1+m^*M^{-1}(p^2-m^\dag m)\cD(p)m^T\right]^{ij} \nn\\
&& \hspace{6cm} \times\,\frac{\pslash_{\al\dot\be}}{p^2} \re^{-\ri p\, (x-y)}  \nn\\
\langle \nu^i_\al (x) N^{j\be} (y) \rangle &=& \ri\int\frac{d^4p}{(2\pi)^4}\left[m^*M^{-1}(p^2- m^\dag m )\cD(p) \right]^{ij} \de_\al^\be \re^{-\ri p\, (x-y)} \nn\\
\langle \nu^i_\al (x) \bN^j_\dbe (y) \rangle &=& \ri\int\frac{d^4p}{(2\pi)^4}\left[m^* \cD(p)^*\right]^{ij} \pslash_{\al\dbe} \re^{-\ri p\, (x-y)}\nn\\
\langle \bnu^{i\dal} (x) \nu^{j\be} (y) \rangle &=& \ri\int\frac{d^4p}{(2\pi)^4}\left[1+m\cD(p)(p^2-m^Tm^*)M^{-1}m^\dag\right]^{ij} \nn\\
&& \hspace{6cm} \times\,\frac{\pslashbar^{\dal\be}}{p^2} \re^{-\ri p\, (x-y)}  \nn\\
\langle \bnu^{i\dal} (x) \bnu^j_\dbe (y) \rangle &=&\ri \int\frac{d^4p}{(2\pi)^4}\left[m \cD(p) m^T\right]^{ij} \de^\dal_\dbe \re^{-\ri p\, (x-y)} \nn\\
\langle \bnu^{i\dal} (x) N^{j\be} (y) \rangle &=& \ri\int\frac{d^4p}{(2\pi)^4}\left[m \cD(p)\right]^{ij} \pslashbar^{\dal\be} \re^{-\ri p\, (x-y)}\nn\\
\langle \bnu^{i\dal} (x) \bN^j_\dbe (y) \rangle &=& \ri\int\frac{d^4p}{(2\pi)^4}\left[m\cD(p)(p^2-m^Tm^*) M^{-1} \right]^{ij} \de^\dal_\dbe \re^{-\ri p\, (x-y)} \nn\\
\langle N^i_\al (x) \nu^{j\be} (y) \rangle &=& \ri\int\frac{d^4p}{(2\pi)^4}\left[\cD(p)(p^2-m^Tm^*)M^{-1} m^\dag\right]^{ij} \de_\al^\be \re^{-\ri p\, (x-y)} \nn\\
\langle N^i_\al (x) \bar\nu^j_\dbe (y) \rangle &=& \ri\int\frac{d^4p}{(2\pi)^4}\left[\cD(p) m^T\right]^{ij} \pslash_{\al\dbe} \re^{-\ri p\, (x-y)}\nn\\
\langle N^i_\al (x) N^{j\be} (y) \rangle &=& \ri\int\frac{d^4p}{(2\pi)^4}\left[ p^2 \cD(p) \right]^{ij} \de_\al^\be \re^{-\ri p\, (x-y)} \nn\\
\langle N^i_\al (x) \bar{N}^j_{\dot\be} (y) \rangle &=& \ri\int\frac{d^4p}{(2\pi)^4}\left[\cD(p) (p^2-m^Tm^*)M^{-1}  \right]^{ij} \pslash_{\al\dot\be} \re^{-\ri p\, (x-y)} \nn \\
\langle \bN^{i\dal} (x) \nu^{j \be} (y) \rangle &=& \ri\int\frac{d^4p}{(2\pi)^4}\left[\cD(p)^* m^\dag\right]^{ij} \pslashbar^{\dal\be} \re^{-\ri p\, (x-y)}\nn\\
\langle \bN^{i\dal} (x) \bnu^j_\dbe (y) \rangle &=& \ri\int\frac{d^4p}{(2\pi)^4}\left[ M^{-1}(p^2-m^\dag m)\cD(p)m^T\right]^{ij} \de^\dal_\dbe \re^{-\ri p\, (x-y)} \nn\\
\langle \bN^{i\dal} (x) N^{j\be} (y) \rangle &=& \ri\int\frac{d^4p}{(2\pi)^4}\left[M^{-1}(p^2-m^\dag m)\cD(p)  \right]^{ij} \pslashbar^{\dal\be} \re^{-\ri p\, (x-y)} \nn \\
\langle \bN^{i\dal} (x) \bN^j_\dbe (y) \rangle &=& \ri\int\frac{d^4p}{(2\pi)^4}\left[ \cD(p)^* p^2 \right]^{ij} \de^\dal_\dbe \re^{-\ri p\, (x-y)} \nn
\eea
where $SL(2,\mathbb{C})$ indices are to be raised and lowered {\em from the left},
as explained in the Appendix.
 The bracket notation $\langle \cdots \rangle\equiv\langle 0|{\rm T}(\cdots)|0\rangle$
 is short-hand for the
time-ordered two-point function. The $\ri\eps$ prescription for $\cD(p)$ (not written out
here) is always such that the analytic continuation to Euclidean propagators works
in the usual way. Note that one cannot simply use hermitian conjugation to check
these expressions, because hermitian conjugation turns a time-ordered product into an
anti-time-ordered product. One can check, however, that all expressions are
consistent with the anti-commutation properties of the fermionic operators.

As a crucial feature of these propagators we note the fall-off properties of the
off-diagonal components at large momenta: unlike the usual Dirac propagator,
these decay like $|p|^3$ or even $|p|^4$, and it is this feature which will account
for the UV finiteness of all the diagrams that we will compute in later sections.
Alternatively, the UV finiteness can also be seen in the `propagator picture' with diagonal propagators from (\ref{MNdiag}): there, the propagators have  the usual fall-off properties, while the softened UV behavior of the diagrams is due to cancellations between different
diagrams arising from the vertices, which are now off-diagonal. Of course, these
cancellations, as well as the final results for the amplitudes, are independent of
specific choices such as (\ref{Mdiag}).

Let us mention one possible application that demonstrates the utility
of the formalism developed here, namely neutrinoless double $\beta$ decay,
see \cite{MSV} for a very recent discussion and bibliography. Inspection of the
relevant diagram (see below) shows that the amplitude for this process
directly `measures'  the propagator components $\langle\nu^i_\al\nu^j_\be\rangle$
and  $\langle \bar{\nu}^i_{\dal} \bar{\nu}^j_{\dbe}\rangle$ listed in
(\ref{NProp}). The $|p|^{-4}$ decay of the
$\langle\nu^i_\al\nu^j_\be\rangle$ and  $\langle \bar{\nu}^i_{\dal} \bar{\nu}^j_{\dbe}\rangle$
propagators for large momenta is different from the $|p|^{-2}$ behavior in models {\em without} right-chiral neutrinos,
where the Majorana mass is induced by a (non-renormalizable) dimension-5 operator
$\sim \, \eps^{\al\be} X_{ij} (\Phi^T\eps L^i_\al) (\Phi^T\eps L^j_\be) + \hc$. In the neutrinoless double beta decay the external momenta are very small, but for larger momenta this
behavior could be used in principle to discriminate between our model, and one
where the left-chiral neutrino is treated as a Majorana particle.

\begin{center}
\includegraphics[width=10cm,viewport= 150 570 500 720,clip]{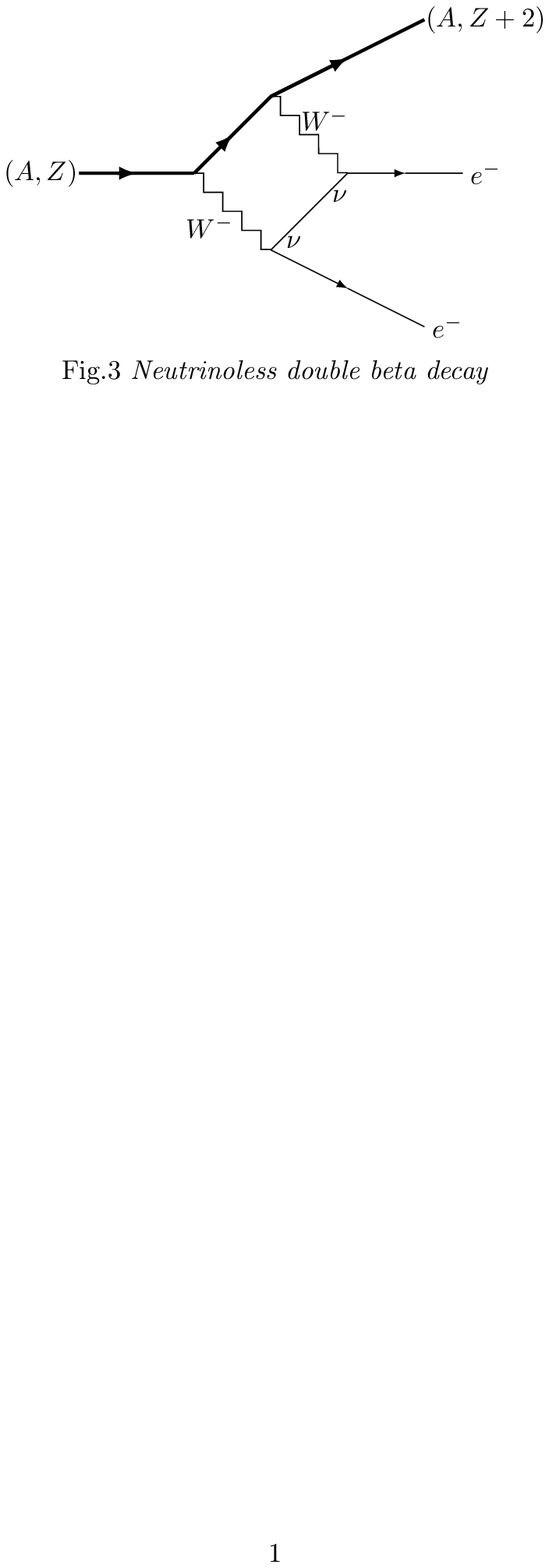} \end{center}

\section{Light vs. heavy neutrinos}

Although our main result does not depend on such choices, it is
occasionally useful to represent the fermion neutrino matrices in a specific
form; we refer readers to \cite{Lindner2,Grimus} for up-to-date discussions of neutrino
masses and mixing. In particular, assuming the real diagonal form  (\ref{Mdiag})
it is not difficult from our formulae to recover the usual seesaw mass formula
\cite{Min,seesaw,Yan,Mo2} from the poles of the propagator matrix $\cD(p)$. More
precisely, the mass eigenvalues are obtained by replacing $p^2$ in
(\ref{defDp}) with the parameter $\lambda$, and demanding
\beq
 \det\Big[ \big( \la - m^Tm^* \big)M^{-1}
    \big( \la - m^\dagger m \big) - M \la \Big]\;=\;  0
\eeq

In this form, the small and large eigenvalues (whose reality follows from the
manifest hermiticity of the mass term in (\ref{KinOp})) are still `entangled'.
For this reason we factorize the total mass matrix by ordering it in
powers of the `small' matrix $m$ as follows,\footnote{We here assume (mainly for
 simplicity) that all eigenvalues of $M^2$ are large in comparison with those of
 $m^\dagger m$; for other mass patterns these arguments may have to be revisited.}
\bea
 &&\!\!\!\!\!\!\!\!\!\det\Big[\left( \la M^{-1} - M-m^T m^* M^{-1}  -M^{-1}m^\dagger m    + M^{-2}m^T m^* M^{-1} m^\dagger m +\cdots\right)\cdot  \nn\\
&& \qquad\qquad\qquad     \cdot\left(\la-M^{-1}m^T m^* M^{-1} m^\dagger m +\cdots\right)\Big] = 0
\eea
where dots stand for higher powers of the small matrix $mM^{-1}$. It is then obvious that, in very good approximation, the determinant equation factorizes into a product of a factor
$\det(\lambda - M^2)$ yielding the large eigenvalues $(M_1^2,M_2^2,M_3^2)$
in (\ref{Mdiag}) for the heavy neutrinos, and a second factor for the light neutrinos.
The latter can be cast in the form
\beq\label{mf}
\det(\la-\mf^\dagger \mf)=0\; ,\ \ \ \mf:= m M^{-1}m^T+\dots  = \mf^T\; ,
\eeq
yielding the mass eigenvalues of the light neutrinos (here we have used the fact
that, for arbitrary square matrices $A$ and $B$, the matrices $AB$ and $BA$ have
the same eigenvalues). The (mass)$^2$ matrix $\mf\mf^\dagger$ thus represents a
matrix generalization of  the usual seesaw formula.

Instead of looking for the poles of $\cD(p)$, we can also arrive at this result by direct consideration of the neutrino mass matrix. The parametrization of mass vs. interaction eigenstates of the light neutrinos is usually given as
\beq\label{U1}
\left(\begin{array}{c} \nu_e\\[1mm] \nu_\mu\\[1mm] \nu_\tau
                                               \end{array}\right)  =
\tilde\cU \left(\begin{array}{c} \nu_1\\[1mm] \nu_2\\[1mm] \nu_3
                                               \end{array}\right)
\eeq
where $\tilde\cU$ is a unitary matrix. In our case this formula is incomplete,
due to the admixture of the heavy neutrinos. To spell out the precise relation between
the 3-by-3 matrix $\tilde\cU$ and the unitary 6-by-6 matrix $\cU$ introduced in
(\ref{U}), we substitute the redefined fields from (\ref{MNdiag}) into (\ref{kinterms}),
demanding the mass term [= second line of (\ref{MNdiag})] to be diagonal in the redefined neutrino fields. Keeping in mind that this redefinition does not mix spinors with dotted and undotted $SL(2,{\mathbb{C}})$ indices, we obtain the condition
\beq\label{UmU}
\cU^* \begin{bmatrix}0&m\\m^T&M\end{bmatrix} \cU^\dag = \begin{bmatrix}m'&0\\0&M'
\end{bmatrix} =\ \cU \begin{bmatrix}0&m^*\\m^\dag&M\end{bmatrix} \cU^T
\eeq
where $m'$ and $M'$ are real diagonal 3-by-3 matrices. This implies
\bea\label{massmatrixsquared}
\begin{bmatrix}m'^2&0\\0&M'^2\end{bmatrix} &=& \cU \begin{bmatrix}0&m^*\\m^\dag&M\end{bmatrix} \cU^T \cU^* \begin{bmatrix}0&m\\m^T&M\end{bmatrix} \cU^\dag = \nn \\[2mm]
&=& \cU \begin{bmatrix}m^*m^T&m^*M\\[1mm]Mm^T&m^\dag m+M^2\end{bmatrix} \cU^\dag
\eea
To relate $\tilde\cU$ to $\cU$, we make the ansatz
\beq
\cU = \begin{bmatrix}
             \cU_1 & \cU_2 \\[1mm]
             \cU_3 & \cU_4
\end{bmatrix}  =
\begin{bmatrix}\tilde{\cU}&0\\0& \tilde{\cU}_M\end{bmatrix}\cR
\eeq
where $\tilde{\cU}$ and $\tilde{\cU}_M$ are both unitary 3-by-3 matrices.
Inserting this ansatz into (\ref{UmU}) and expanding in powers of the `small'
matrix $mM^{-1}$ up to second order, we obtain
\beq
\cR=\begin{bmatrix} {\bf 1}-\frac12 m^*M^{-2}m^T &-m^*M^{-1}  \\[1mm]
M^{-1}m^T &{\bf 1} -\frac12 M^{-1}m^Tm^*M^{-1}\end{bmatrix}+O\left(\left(mM^{-1}\right)^3\right)
\eeq
and, up to diagonal phase redefinitions, the matrices $\tilde{\cU}$ and $\tilde{\cU}_M$  are determined by the conditions
\bea
m' &=& -\tilde{\cU}^* mM^{-1}m^T \tilde{\cU}^\dagger
    \equiv -\tilde{\cU}^* \mf  \, \tilde{\cU}^\dagger \nn\\
M'&=& \tilde{\cU}_M^*\left(M+\frac12  m^Tm^*M^{-1}+\frac12M^{-1}m^\dagger m\right)
 \tilde{\cU}_M^\dagger
\eea
Multiplying these matrices by the complex conjugate matrices and using the
reality of $m'$ and $M'$, we see again that $\tilde\cU$ diagonalizes the generalized
seesaw  (mass)$^2$ matrix $\mf^\dagger\mf$. In other words, we have now rather explicit
expressions for the eigenvalues of mass matrices of light ($m'$) and heavy ($M'$)
neutrinos, as well as for the unitary matrix relating the `propagator picture' and
the `vertex picture'. In lowest approximation the matrix $\tilde{\cU}$ in (\ref{U1}) is equal to $\cU_1$ in (\ref{U}).

Obviously, the mixing of light and heavy neutrinos has several interesting physical
implications. For instance, the heaviest among the light neutrinos can decay
into the lightest neutrino via the emission of `soft axions'. This process is possible because in our scenario the axion is expected to be almost massless [for instance, with a mass $m_{\rm axion} = \cO( 10^{-8} \, {\rm eV})$], and thus much lighter than even the lightest neutrino.

\section{$aZ$ mixing}

We are here concerned with the effective couplings of the `invisible axion' $a(x)$ to `visible' SM fields. Such couplings must arise through loop diagrams (with the exception of couplings to neutrinos through the mixing described in the previous section), as there are no direct couplings at the tree level, a fact which according to our proposal can explain the extreme smallness of the axion couplings to standard matter. Assuming $\vev \neq 0$ we now fix the free normalization parameter in (\ref{phi}) once and for all to the value $\mu = \vev$, as in (\ref{phikin}), in order to obtain the canonical normalization for the kinetic term of $a(x)$ in the classical lagrangian.

However, before we proceed to the actual computation we need to discuss the mixing
of $a$ with $Z$ bosons. The result described in this section shows that in principle, there can be large effects (in this case at one loop). As we will see, these  fail to contribute to the axion-gluon coupling only because of the vanishing chiral anomaly of the SM, but they can nevertheless dominate in other processes. There is a similar mixing between the $Z$ boson and the scalar $\vp'(x)$, as well as a mixing between $a(x)$ and the
standard Higgs boson, but these couplings turn out to be suppressed by an extra
factor of the light neutrino masses, and can thus be neglected.

The mixing of axions with gauge bosons, which may arise at one loop or higher
loop orders, may in principle occur between $a(x)$ and any {\em neutral} gauge field $\cA_\mu(x)$ at the quadratic level, leading to extra terms $\propto \partial_\mu a \,\cA^\mu$
in the effective Lagrangian. However, gauge invariance immediately forbids such couplings if the gauge symmetry is unbroken, whence $a(x)$ cannot couple in this way to either photons or gluons.  For the broken sector, this argument does not hold, and by charge conservation, we are therefore left with possible quadratic couplings of $a$ to the $Z$ boson,
\beq\label{mix}
\cL_{\rm mix} = \eps \partial_\mu a Z^\mu
\eeq
where the parameter $\eps$ is of dimension one.

We now determine the mixing coefficient $\eps$ at one loop (with right-chiral neutrinos in the loop), and show that $\eps$ is proportional to the sum of the light neutrino masses.
The relevant diagram is shown below, and involves the off-diagonal
$\langle N\nu\rangle$ and $\langle N\bar{\nu}\rangle$ ($\langle \bN\nu\rangle$ and $\langle \bN\bar{\nu}\rangle$ in the second amplitude) components of the
neutrino propagators  in the loop.

\begin{center}
\includegraphics[width=7cm,viewport= 150 600 350 720,clip]{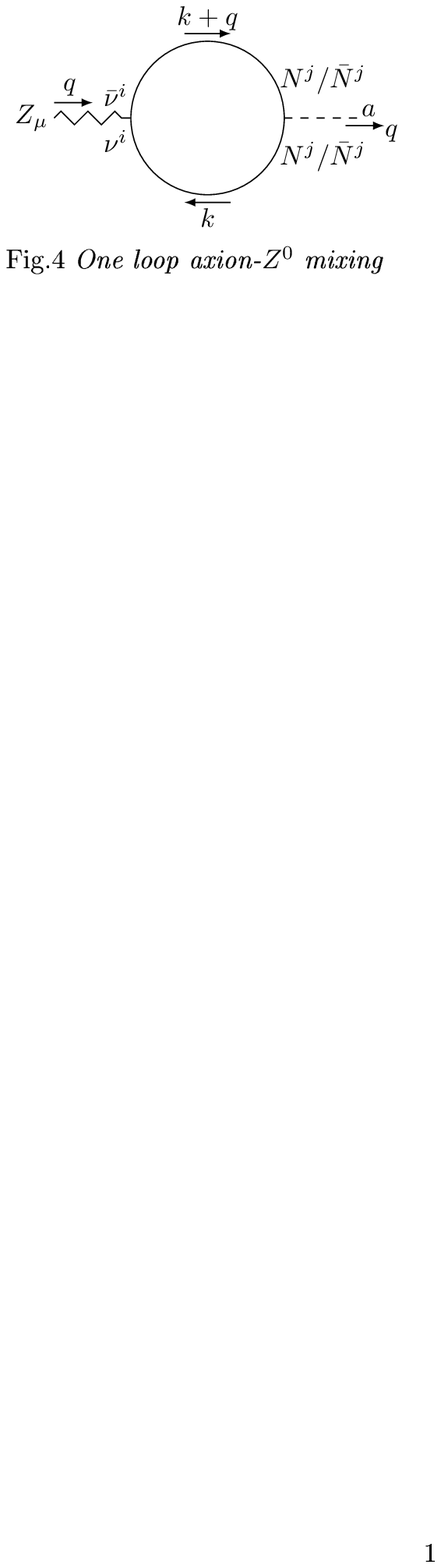}
\end{center}

This diagram gives rise to the following Feynman integral which can be evaluated in the standard fashion; note that, while naive power counting would suggest the presence of divergences, {\em this diagram is actually UV finite} because of the fast decay of the propagator components $\langle N\nu\rangle$, $\langle N \bar\nu\rangle$, $\langle \bN\nu\rangle$ and $\langle \bN \bar\nu\rangle$ at large momenta.
\bea
-\ri\cM^{\mu}_{aZ}(q) &=& \nn \\[1mm]
&& \!\!\!\!\!\!\!\!\!\!\!\!\!\!\!\!\!\!\!\!\!\!\!\!\!\!\!\!\!\!=
(-1) \sum_{i,j}\int\frac{\rd^4k}{(2\pi)^4} \left\{ \left(-\ri\frac{g_2}{2\cos\th_W} \bar\si^{\mu\dal_1\be_1} \right) \times \right.\nn \\
&&  \!\!\!\!\!\!\!\!\!\!\!\!\!\!\!\!\!\!\!\!\!\!\!\! \qquad
\times \left[ \langle \nu^i_{\be_1}N^{j\al_2} (k) \rangle \left( \frac{M_j}{\sqrt{2}\vev} \right) \langle N^j_{\al_2}\bnu^i_{\dal_1} (k+q) \rangle \right.  \nn \\
&&  \left.\left.\!\!\!\!\!\!\!\!\!\!\!\!\!\!\!\!\!\!\!\!\!\!\!\!
\qquad\qquad + \langle \nu^i_{\be_1}\bar{N}^{j}_{\dal_2} (k) \rangle \left( \frac{-M_j}{\sqrt{2}\vev} \right)
\langle \bar{N}^{j\dal_2}\bnu^i_{\dal_1} (k+q) \rangle \right]\right\} \nn \\
&&  \!\!\!\!\!\!\!\!\!\!\!\!\!\!\!\!\!\!\!\!\!\!\!\! =\;
(-1) \sum_{i,j}\int\frac{\rd^4k}{(2\pi)^4} \left\{ \left(-\ri\frac{g_2}{2\cos\th_W} \bar\si^{\mu\dal_1\be_1} \right) \times \right.\nn \\
&&  \!\!\!\!\!\!\!\!\!\!\!\!\!\!\!\!\!\!\!\!\!\!\!\!
\times \left[ \left(\ri[m^*\cD(k)^*(k^2-m^\dag m)M^{-1}]^{ij}\de^{\al_2}_{\be_1}\right) \left(\frac{M_j}{\sqrt{2}\vev} \right)\cdot\right.\\
&&\left.
\cdot\left(\ri (\kslash+\qslash)_{\al_2\dal_1} [\cD(k+q) m^T]^{ji} \right) \right. + \nn \\
&& +  \left(\ri [m^*\cD(k)^*]^{ij} \kslash_{\be_1\dal_2}\right) \left( \frac{-M_j}{\sqrt{2}\vev} \right)\cdot\nn\\ &&\left.\left.\ \ \ \cdot\left(\ri[M^{-1}((k+q)^2-m^\dag m)\cD(k+q)m^T]^{ji}\de^{\dal_2}_{\dal_1}\right) \right]\right\} = \nn \\[1mm]
&&  \!\!\!\!\!\!\!\!\!\!\!\!\!\!\!\!\!\!\!\!\!\!\!\!\!\!\!\!\!\!\!\!\!\!\!\!\!\!\!\! =
-\ri\frac{g_2}{2\sqrt{2}\vev\cos\th_W }\sum_{i}\int\frac{\rd^4k}{(2\pi)^4} \,
{\rm Tr}\,  \big( \bar\si^\mu \si^\nu \big)
 \times \nn \\
&& \!\!\!\!\!\!\!\!\!\!\!\!\!\!\!\!\!\!\!\!\!\!
 \times \Big[m^* \cD(k)^* [(k^2-m^\dag m)(k+q)_\nu - k_\nu ((k+q)^2-m^\dag m)] \cD(k+q) m^T\Big]^{ii} = \nonumber \\[1mm]
&&  \!\!\!\!\!\!\!\!\!\!\!\!\!\!\!\!\!\!\!\!\!\!\!\!\!\!\!\!\!\!\!\!\!\!\!\!\!\!\!\! =
-\ri\frac{g_2}{\sqrt{2}\vev\cos\th_W } \int\frac{\rd^4k}{(2\pi)^4} \text{Tr}\Big[m^* \cD(k)^* [q^\mu (k^2-m^\dag m)- k^\mu (2kq+q^2)] \cD(k+q) m^T \Big] \nn \\[1mm]
&&\!\!\!\!\!\!\!\!\!\!\!\!\!\!\!\!\!\!\!\!\!\!\!\!\!\!\!\!\!\!\!\!\!\!\!\!\!\approx
 \ri\frac{g_2}{\sqrt{2}\vev\cos\th_W } \sum_i (m^T m^*)^{ii} \int\frac{\rd^4k}{(2\pi)^4} \frac{[k^\mu (2kq+q^2)- q^\mu k^2]M_i^2}{k^2(k^2-M_i^2)(k+q)^2[(k+q)^2-M_i^2]}\nn
\eea
We are here interested only in the result for small axion momentum $q^\mu$,
in which case the integral can be worked out to be
\bea
 -\ri\cM^{\mu}_{aZ}(q) &\approx&
 \ri q_\nu \frac{g_2}{\sqrt{2}\vev\cos\th_W } \sum_i (m^T m^*)^{ii} \int\frac{\rd^4k}{(2\pi)^4} \frac{(2k^\mu k^\nu- \eta^{\mu\nu} k^2)M_i^2}{[k^2(k^2-M_i^2)]^2} = \nonumber \\
&=& -q^\mu \frac{g_2}{(4\pi)^2 2\sqrt{2}\vev\cos\th_W } \sum_{i,j} |m^{ij}|^2
\eea
Remembering (\ref{mM}) and making the (reasonable) assumption $Y_M\sim\cO(1)$
we see that
\beq
\sum_{i,j} \frac{|m^{ij}|^2}{\vev} \, \sim\, \sum_{i,j} \frac{|m^{ij}|^2}{M} \sim \, \sum m_\nu
\eeq
whence we arrive at the claimed result $\eps \sim \sum m_\nu$, that is, the
mixing is proportional to the sum of the light neutrino masses. For the mixing
between $Z^\mu$ and the scalar $\vp'$ the two contributions to the above integral
would appear with opposite signs, leading to a further cancellation, with a
mixing parameter of order $\cO(m_\nu^2)$ (which hence can be ignored).

The $aZ$ mixing described above can lead, via subsequent $Z$ couplings, to
further couplings of the axion to other SM fields like quarks and leptons. When considering
physical effects such as axion cooling in stars from axion emission from
leptons or quarks, these contributions must be taken into account. For the same
reason, one would also expect the above mixing to  contribute to the effective
coupling of axions to gluons via the anomaly diagram
\begin{center}
\includegraphics[width=10cm,viewport= 150 630 470 720,clip]{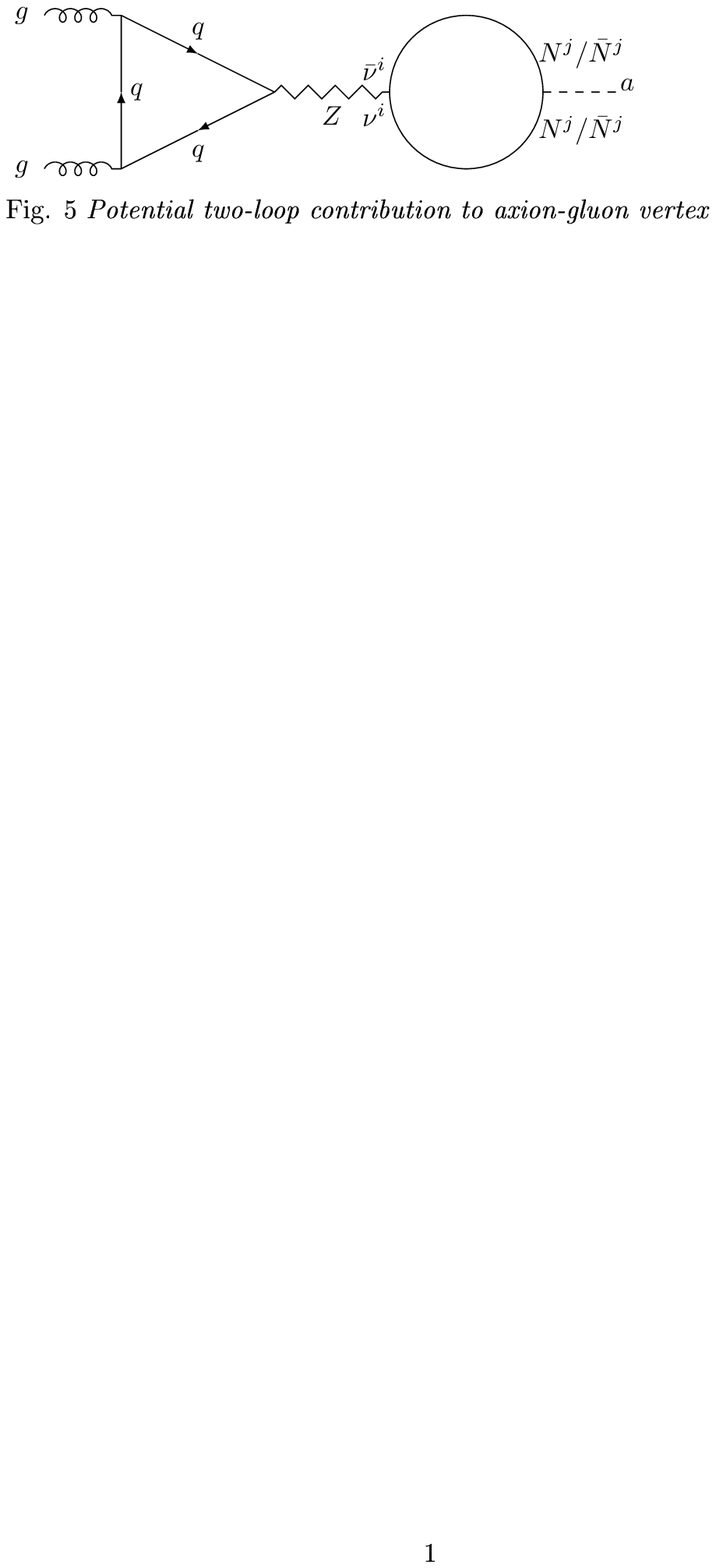} \end{center}
which, if non-vanishing, would be more important than the three-loop computation
we are going to perform in the final section of this paper! However, here we meet
a pleasant surprise, which will simplify our subsequent considerations substantially:
{\em when summing over the quarks in this diagram one obtains a
vanishing result because of the cancellation of all gauge anomalies in the SM}.
In other words, when determining the effective coupling of the axion to gluons we
can ignore the above mixing (and similar diagrams). The same conclusion
holds for the couplings of the axion to photons, when all SM fermions are
summed over. The important fact is therefore that the
non-vanishing effective couplings arise solely via the three-loop diagrams
with an attached neutrino triangle which we will work out in the remaining sections.

\section{The $aW^+W^-$ triangle at one loop}

Our main proposal relies essentially on a new effect producing an `anomaly-like'
amplitude from a triangle diagram involving neutrino triangles, which in turn
gives rise to the effective coupling of $a(x)$ to $W$-bosons; the relevant diagram
is  depicted in the figure below. The similarity of this triangle diagram with the
well known one producing the triangle anomaly is obvious, yet the anomaly-like effect here is {\em not} due to a linear UV divergence, but rather to the mixing of the neutrino components. Technically speaking, one reason for this is that the neutrino propagators involve $\si$-matrices rather than $\ga$-matrices, and the trace
\beq
\label{t}
t^{\mu\nu\rho\lambda} \equiv {\rm Tr}\, \bsi^\mu\si^\nu\bsi^\rho\si^\lambda
= 2\big(\eta^{\mu\nu}\eta^{\rho\lambda} - \eta^{\mu\rho}\eta^{\nu\lambda}
     + \eta^{\mu\lambda} \eta^{\nu\rho}  +  \ri\eps^{\mu\nu\rho\lambda}\big)
\eeq
over $\si$-matrices generates both parity even as well as parity odd terms, unlike the corresponding $\ga$-matrix trace ${\rm Tr}\, \ga^\mu\ga^\nu \ga^\rho \ga^\lambda$.

In this section we take the first step towards our final goal of determining the effective axion-gluon coupling by calculating the $aW^+W^-$ vertex; this result will then be used as an input in the calculation of the axion-quark diagram in the following section, which in turn will be substituted in the final step into the three-loop diagram yielding the effective axion-gluon vertex. Once more, we emphasize that {\em all diagrams
in this and the following sections are UV finite}, even though
naive power counting might suggest otherwise. Let us also point out that
there are similar diagrams with two external $Z$-bosons, where the triangle
is `purely neutrino'  (that is, all internal lines are neutrino propagators).
In accordance with our basic strategy for computing effective couplings
outlined in the introduction, we will however disregard these diagrams, because
they necessarily contain a light neutrino propagator ($\langle\nu\nu\rangle$ or
$\langle\nu\bar\nu\rangle$ or $\langle\bar\nu\bar\nu\rangle$) on the internal line
connecting the two $Z$-boson vertices.

\begin{center}
\includegraphics[width=8cm,viewport= 170 620 420 720,clip]{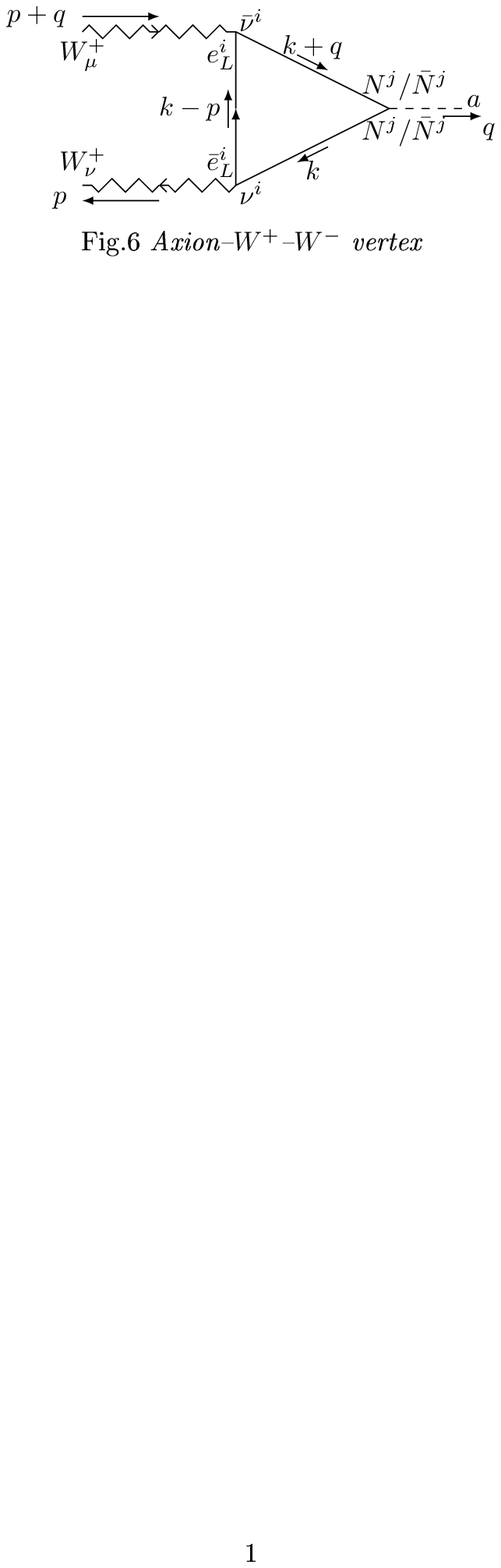} \end{center}
\noindent
The above Feynman diagram corresponds to the integral

\bea
&&\!\!\!\!\!\!\!\!\!\!\!\!\!\!\!
 -\ri\cM^{\mu\nu}_{aWW}(p,q) = \nn\\[1mm]
&=& (-1) \sum_{i,j}\int\frac{\rd^4k}{(2\pi)^4} \left\{ \left(-\ri\frac{g_2}{\sqrt{2}} \bar\si^{\mu\dal_1\be_1} \right) \langle e^i_{L\be_1}\bar{e}^i_{L\dal_2} (k-p) \rangle \left(-\ri\frac{g_2}{\sqrt{2}} \bar\si^{\nu\dal_2\be_2} \right) \times \right.\nn\\
&& \times \left[ \langle \nu^i_{\be_2}N^{j\al_3} (k) \rangle \left( \frac{M_j}{\sqrt{2}\vev} \right) \langle N^j_{\al_3}\bnu^i_{\dal_1} (k+q) \rangle \right.  \nn \\
&&  \left.\left.\!\!\!\!\!\!\!\!\!\!\!\!\!\!\!\!\!\!\!\!\!\!\!\!
\qquad\qquad + \langle \nu^i_{\be_2}\bar{N}^{j}_{\dal_3} (k) \rangle \left( \frac{-M_j}{\sqrt{2}\vev} \right)
\langle \bar{N}^{j\dal_3}\bnu^i_{\dal_1} (k+q) \rangle \right]\right\} = \nn
\eea
\bea
&=& (-1) \sum_{i,j}\int\frac{\rd^4k}{(2\pi)^4} \left\{ \left(-\ri\frac{g_2}{\sqrt{2}} \bar\si^{\mu\dal_1\be_1} \right) \left(\ri\frac{(\kslash-\pslash)_{\be_1\dal_2}}{(k-p)^2-m_{e_i}^2}\right) \left(-\ri\frac{g_2}{\sqrt{2}} \bar\si^{\nu\dal_2\be_2} \right) \times \right.\nn\\
&& \times \left[ \left(\ri[m^*\cD(k)^*(k^2-m^\dag m)M^{-1}]^{ij}\de^{\al_3}_{\be_2}\right) \left(\frac{M_j}{\sqrt{2}\vev} \right)\cdot\right.\\
&&\left.
\cdot\left(\ri (\kslash+\qslash)_{\al_3\dal_1} [\cD(k+q) m^T]^{ji} \right) \right. + \nn \\
&& +  \left(\ri [m^*\cD(k)^*]^{ij} \kslash_{\be_2\dal_3}\right) \left( \frac{-M_j}{\sqrt{2}\vev} \right)\cdot\nn\\ &&\left.\left.\ \ \ \cdot\left(\ri[M^{-1}((k+q)^2-m^\dag m)\cD(k+q)m^T]^{ji}\de^{\dal_3}_{\dal_1}\right) \right]\right\} = \nn\\[1mm]
&=& -\ri\frac{g_2^2}{2\sqrt{2}\vev }\sum_{i}\int\frac{\rd^4k}{(2\pi)^4}  \frac{(k-p)_\la}{(k-p)^2-\me^2} \left( (\bar\si^\mu)^{\dal_1\be_1} \si^\la_{\be_1\dal_2} (\bar\si^\nu)^{\dal_2\be_2} \si^\rho_{\be_2\dal_1} \right) \nn \\
&& \!\!\!\!\!\!\!\times \Big[m^* \cD(k)^* [(k^2-m^\dag m)(k+q)_\rho - k_\rho ((k+q)^2-m^\dag m)] \cD(k+q) m^T\Big]^{ii} = \nn\\[1mm]
&=& \ri\frac{g_2^2}{2\sqrt{2}\vev}\sum_{i}\int\frac{\rd^4k}{(2\pi)^4}  \frac{(k-p)_\la}{(k-p)^2-\me^2} \text{Tr}\{\bar\si^\mu\si^\la\bar\si^\nu\si^\rho\}  \\
&& \times \Big[m^* \cD(k)^*  [k_\rho (2kq+q^2)- q_\rho (k^2-m^Tm^*)] \cD(k+q)m^T \Big]^{ii}\nn
\eea
As before, we are here interested in the leading terms for small axion momentum $q^\mu$ and small $m$, in a basis where $M$ is real diagonal; in this approximation the integral simplifies to
\bea
&&\!\!\!\!\!\!\!\!\!\!\!\!\!\!\!\!\!\!\!\!\!
-\ri\cM^{\mu\nu}_{aWW}(p,q) \approx \ri q^\tau \frac{g_2^2}{2\sqrt{2}\vev }
\sum_{i,j} |m^{ij}|^2\times \text{Tr}\{\bar\si^\mu\si^\la\bar\si^\nu\si^\rho\} \times \nn\\
&&\qquad\qquad\times\,  \int\frac{\rd^4k}{(2\pi)^4}  \frac{(k-p)_\la}{(k-p)^2-\me^2} \frac{(2k_\rho k_\tau- \eta_{\rho\tau} k^2) M_j^2}{[k^2(k^2-M_j^2)]^2}
\eea
This integral can be evaluated by means of  Feynman parameters, with the result
\bea
&&\!\!\!\!\!\!\!\!\!\!\!\!\!\!\!\!\!\!\!\!\!\!\!\!
 -\ri\cM^{\mu\nu}_{aWW}(p,q)  \approx \nn\\[1mm]
&\approx& -q^\tau \frac{g_2^2}{32\pi^2\sqrt{2}\vev }\sum_{i,j} |m^{ij}|^2 M_j^2 \;
 t^{\mu\lambda\nu\rho}  \int_0^1 \rd y\int_0^1 \rd x\, x(1-x)y^3 \nn\\
&& \times \left[ \frac{[(-1+2y) p_\la \eta_{\tau\rho} +(1-y)(p_\tau \eta_{\rho\la}+p_\rho \eta_{\tau\la})]}{[-y(1-y)p^2+(1-y)\me^2+yxM_j^2]^2} \right.  \;  \nn\\
&&\qquad\qquad\qquad \left.+\;\frac{2y(1-y)^2 p_\la (-p^2\eta_{\tau\rho} + 2p_\tau p_\rho)}{[-y(1-y)p^2+(1-y)\me^2+yxM_j^2]^3}\right]
\eea
where we made use of the definition (\ref{t})
(note that the denominator in the integrand will become positive definite after
Wick rotation to Euclidean momenta $p^\mu$). Although we do need the full
expression (for large $p^\mu$) below, it is nevertheless
instructive to specialize this result to small $p$ to get
\bea
&& \!\!\!\!\!\!\!\!\!\!\!\!\!\!\!\!\!\!\!
-\ri\cM^{\mu\nu}_{aWW}(p,q) \approx \nn\\[1mm]
&\approx& -\frac{g_2^2}{32\pi^2\sqrt{2}\vev }\sum_{i,j} |m^{ij}|^2 M_j^2\;
t^{\mu\lambda\nu\rho} q^\tau  \times \\
&& \times \int_0^1 \rd y\int_0^1 \rd x\, \frac{x(1-x)y^3[(-1+2y) p_\la \eta_{\tau\rho} +(1-y)(p_\tau \eta_{\rho\la}+p_\rho \eta_{\tau\la})]}{[(1-y)\me^2+yxM_j^2]^2} = \nn\\[1mm]
&=& -\frac{\al_2}{8\pi\sqrt{2}\vev } \, t^{\mu\lambda\nu\rho} q^\tau \sum_{i,j} |m^{ij}|^2\nn\\
&& \times\left[p_\la \eta_{\tau\rho} \frac{M_j^2(M_j^2-5\me^2)\log\frac{M_j^2}{\me^2}+(M_j^2+3\me^2)(M_j^2-\me^2)}{6(M_j^2-\me^2)^3} \right.  \nn\\
&& \quad \left. + (p_\tau \eta_{\rho\la}+p_\rho \eta_{\tau\la})  \frac{M_j^2(M_j^2+\me^2)\log\frac{M_j^2}{\me^2} - 2M_j^2(M_j^2-\me^2)}{6(M_j^2-\me^2)^3} \right]\nn
\eea
The singularity in this expression for $\me^2=M_j^2$ is spurious.

Let us pause to put this result in perspective. The trace over $\sigma$-matrices
contains both parity even as well as parity odd terms.
The former lead to non-gauge invariant contributions for the amplitude,
proportional to $q^\mu p^\nu + q^\nu p^\mu$ and $(q^\rho p_\rho)\eta^{\mu\nu}$,
respectively. As we already pointed out, such contributions are to be expected because the
electroweak symmetry $SU(2)_w\times U(1)_Y$ is broken. In addition we get a gauge
invariant anomaly-like term $\propto \eps^{\mu\nu\rho\sigma} p_\rho q_\sigma$.
The presence of both these terms is in accord with the fact that parity is maximally
violated in the SM. By contrast, for gauge bosons associated with an unbroken
gauge invariance on the external lines, only the gauge invariant anomaly-like
contribution can survive by the general arguments given in section~2.
Consequently, insertion of the above triangle as a subdiagram
into a higher loop diagram with external photons or gluons will yield only the
gauge invariant anomaly-like amplitude, as we shall explicitly verify.
In particular, for the axion-gluon amplitude we can anticipate that the result will be proportional to $\ri \eps^{\mu\nu\la\si}q_\la p_\si$, as $SU(3_c)$ remains unbroken.
This is the core effect which justifies our claim as to the
emergence of the effective coupling (\ref{Lagg}). In the following sections we will
verify this claim by explicit computation.

\section{Axion-quark diagrams at two loops}

The next step is the the calculation of the axion-quark diagram, which is given
by the following two-loop diagrams:
\begin{center}
\includegraphics[width=7.5cm,viewport= 190 500 420 720,clip]{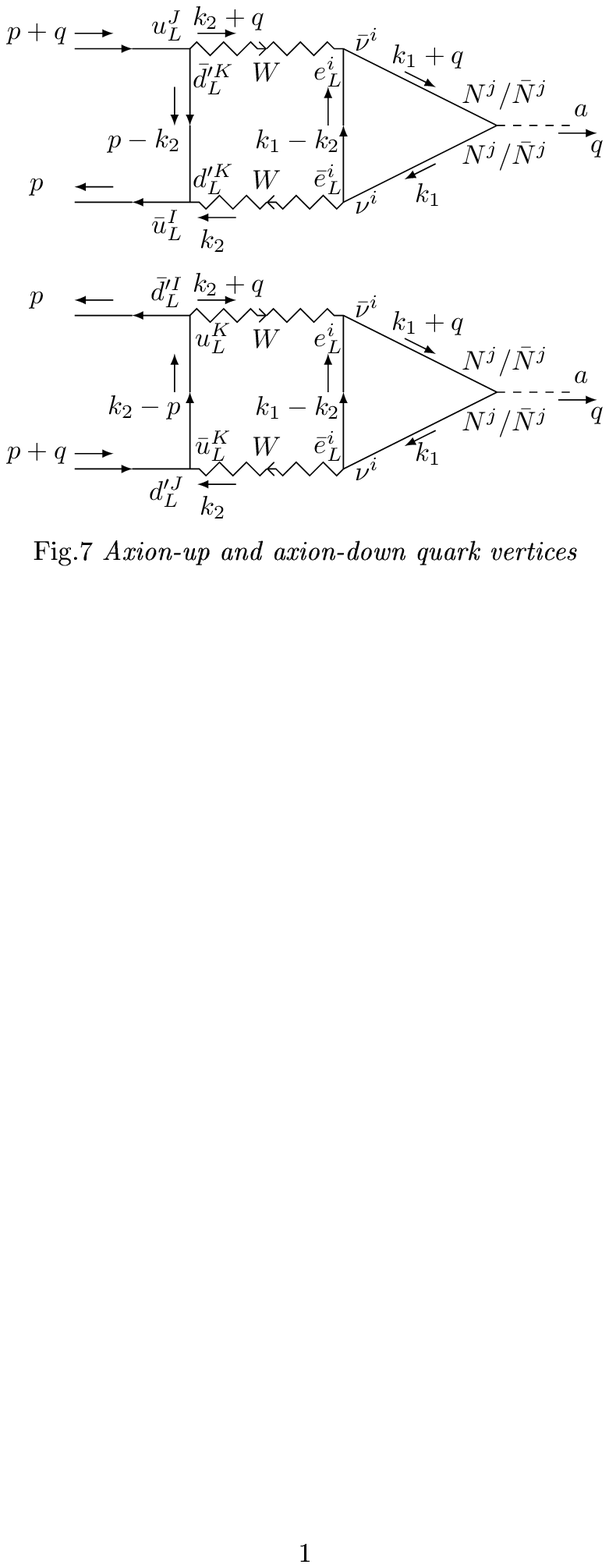} \end{center}
Since this diagram involves the gauge boson propagators on the internal lines we have to
specify the gauge. In the so called $R_\xi$ gauge the propagator reads
\beq
\langle W^{+\mu} W^{-\nu}\rangle (k) =  \frac{1}{k^2-M_W^2}\left(
\eta^{\mu\nu}+(\xi-1)\frac{k^\mu k^\nu}{k^2-\xi M_W^2}\right)
\eeq
In the full calculation one must also include the diagrams with charged Goldstone bosons from the Higgs doublet whose propagator behaves as $1/(k^2-\xi M_W^2)$, and whose contribution vanishes only in the limit $\xi\to\infty$. The calculation then gets very involved even by comparison with the formulae of this and subsequent sections (the full calculation and all details will be given in \cite{LThesis}). The estimate shows that the contribution of these terms is of the same order of magnitude as what we calculate because of the large Yukawa coupling of the top quark. Therefore the final result that we get by neglecting both $k^\mu k^\nu$ parts of the propagators and the contributions from charged Higgs particles is only an estimate of the value of the actual result.

To make the formulae more transparent, we will  use {\em capital indices
$I,J,\dots$ for the quark flavors} in the remainder, and now also write out the
explicit sums over them. With this convention, the above diagrams correspond
to the following Feynman integrals
\bea
&&\!\!\!\!\!\!\!\!\!\!\!\!\!\!\!\!\!
 -\ri\cM_{auu}^{IJ}(p,q) = \nn\\[1mm]
&=& \sum_K \int\frac{\rd^4k_2}{(2\pi)^4} \left(-\ri\frac{g_2}{\sqrt{2}} V^{IK}\bar\si_\nu \right)  \frac{\ri(\pslash-\kslash_2)}{(p-k_2)^2-\md^2} \left(-\ri\frac{ g_2}{\sqrt{2}} (V^\dag)^{KJ}\bar\si_\mu \right)\nn\\
&& \times \frac{\ri}{(k_2+q)^2-M_W^2}\frac{\ri}{k_2^2-M_W^2} (-\ri\cM^{\mu\nu}_{aWW}(k_2,q)) = \nn\\[1mm]
&=& \ri\frac{g_2^2}{2}\sum_K V^{IK}(V^\dag)^{KJ} \int\frac{\rd^4k_2}{(2\pi)^4} \frac{(p-k_2)_\tau}{(p-k_2)^2-\md^2}\frac{1}{(k_2+q)^2-M_W^2}\frac{1}{k_2^2-M_W^2} \nn\\
&& \qquad \times \; \bar\si^\nu \si^\tau \bar\si^\mu  \times (-\ri\cM^{\mu\nu}_{aWW}(k_2,q))
\eea
and
\bea
&& \!\!\!\!\!\!\!\!\!\!\!\!\!\!\!\!\!
-\ri \cM_{add}^{IJ} (p,q) = \nn\\[1mm]
&=& \sum_K \int\frac{\rd^{4}k_2}{(2\pi)^4} \left(-\ri\frac{ g_2}{\sqrt{2}} (V^\dag)^{IK}\bar\si_\mu \right) \frac{\ri(\kslash_2-\pslash)}{(k_2-p)^2-m_{u_K}^2} \left(-\ri \frac{g_2}{\sqrt{2}} V^{KJ} \bar\si_\nu \right)\nn\\
&& \times \frac{\ri}{(k_2+q)^2-M_W^2}\frac{\ri}{k_2^2-M_W^2} (-\ri\cM^{\mu\nu}_{aWW}(k_2,q)) = \nn\\[1mm]
&=& \ri\frac{g_2^2}{2}\sum_K (V^\dag)^{IK}V^{KJ} \int\frac{\rd^{4}k_2}{(2\pi)^4} \frac{(k_2-p)_\tau}{(p-k_2)^2-m_{u_K}^2}\frac{1}{(k_2+q)^2-M_W^2}\frac{1}{k_2^2-M_W^2} \nn\\
&& \qquad \times \; \bar\si^\mu \si^\tau \bar\si^\nu \times (-\ri\cM^{\mu\nu}_{aWW}(k_2,q))
\eea
Using the $\cM^{\mu\nu}_{aWW}$ result already calculated before together with
\bea
\bar\si^\nu \si^\tau \bar\si^\mu \times t^{\mu\lambda\nu\rho} &=& 8\de^\la_\tau \bar\si^\rho
\label{gammarelations1}
\eea
we get, as always for small $q^\mu$,
\bea\label{aqInt1}
&& \!\!\!\!\!\!\!\!\!\!\!\!\!\!\!\!\!
-\ri\cM_{auu}^{IJ}(p,q) = \nn\\[1mm]
&=& -\ri q^\rho\bar\si^\si  \frac{g_2^4}{8\pi^2\sqrt{2}\vev }\sum_K V^{IK}(V^\dag)^{KJ} \sum_{ij} |m^{ij}|^2 M_j^2 \int_0^1 \rd y\int_0^1 \rd x\, x(1-x)y^3 \times \nn\\
&& \qquad
\times \int\frac{\rd^{4}k}{(2\pi)^4} \frac{-(k-p)^\la}{(k-p)^2-\md^2}\frac{1}{(k^2-M_W^2)^2} \times \\
&& \qquad \qquad\times \left[ \frac{[(-1+2y) k_\la \eta_{\rho\si} +(1-y)(k_\rho \eta_{\si\la}+k_\si \eta_{\rho\la})]}{[-y(1-y)k^2+(1-y)\me^2+yxM_j^2]^2}\right. \nn\\
&&\qquad\qquad\qquad\qquad
 \left.+\; \frac{2y(1-y)^2 k_\la (-k^2\eta_{\rho\si} + 2k_\rho k_\si)}{[-y(1-y)k^2+(1-y)\me^2+yxM_j^2]^3}\right]\nn
\eea
and
\bea\label{aqInt2}
&& \!\!\!\!\!\!\!\!\!\!\!\!\!\!\!\!\!
-\ri\cM_{add}^{IJ}(p,q) = \\
&=& -\ri q^\rho\bar\si^\la  \frac{g_2^4}{8\pi^2\sqrt{2}\vev }\sum_K  (V^\dag)^{IK}V^{KJ} \sum_{ij} |m^{ij}|^2 M_j^2 \int_0^1 \rd y\int_0^1 \rd x\, x(1-x)y^3 \times\nn\\
&& \times \int\frac{\rd^{4}k}{(2\pi)^4} \frac{(k-p)^\si}{(k-p)^2-m_{u_K}^2}\frac{1}{(k^2-M_W^2)^2} \times \nn\\
&& \qquad\times \left[ \frac{[(-1+2y) k_\la \eta_{\rho\si} +(1-y)(k_\rho \eta_{\si\la}+k_\si \eta_{\rho\la})]}{[-y(1-y)k^2+(1-y)\me^2+yxM_j^2]^2} \right.\nn\\
&&\qquad\qquad\qquad \left.+\; \frac{2y(1-y)^2 k_\la (-k^2\eta_{\rho\si} + 2k_\rho k_\si)}{[-y(1-y)k^2+(1-y)\me^2+yxM_j^2]^3}\right] \nn
\eea
Again we employ Feynman parameters to obtain
\bea
&& \!\!\!\!\!\!\!\!\!\!\!\!\!\!\!\!\!
-\ri\cM_{auu}^{IJ}(p,q) = \nn\\[1mm]
&=& q^\rho \bar\si^\si \frac{g_2^4}{64\sqrt{2}\pi^4\vev} \sum_{i,j,K} V^{IK}(V^\dag)^{KJ}|m^{ij}|^2M_j^2\int_0^1 \rd x\int_0^1 \rd y\int_0^1 \rd z\int_0^1 \rd t \nn\\
&& \times \bigg\{\frac{f_1^u(x,y,z,t) \eta_{\rho\si}}{[\MM^2_{ijK}(x,y,z,t;p)]^2} + \frac{f_2^u(x,y,z,t)p^2 \eta_{\rho\si} + f_3^u(x,y,z,t) p_\rho p_\si}{[\MM^2_{ijK}(x,y,z,t;p)]^3}  \nn\\
&&\qquad\qquad\qquad
+\;\frac{f_4^u(x,y,z,t)p^4\eta_{\rho\si} + f_5^u(x,y,z,t)p^2 p_\rho p_\si}{[\MM^2_{ijK}(x,y,z,t;p)]^4}
\bigg\}
\eea
and
\bea
&& \!\!\!\!\!\!\!\!\!\!\!\!\!\!\!\!\!
-\ri\cM_{add}^{IJ}(p,q) = \nn\\
&=& q^\rho \bar\si^\si \frac{g_2^4 }{64\sqrt{2}\pi^4\vev} \sum_{i,j,K} (V^\dag)^{IK}V^{KJ}|m^{ij}|^2M_j^2 \int_0^1 \rd x\int_0^1 \rd y\int_0^1 \rd z\int_0^1 \rd t \nn\\
&& \times \bigg\{\frac{f_1^d(x,y,z,t) \eta_{\rho\si}}{[\widetilde{\MM^2}_{ijK}(x,y,z,t;p)]^2} + \frac{f_2^d(x,y,z,t)p^2 \eta_{\rho\si} + f_3^d(x,y,z,t) p_\rho p_\si}{[\widetilde{\MM^2}_{ijK}(x,y,z,t;p)]^3}  \nn\\
&&\qquad\qquad\qquad
+\;\frac{f_4^d(x,y,z,t)p^4\eta_{\rho\si} + f_5^d(x,y,z,t)p^2 p_\rho p_\si}{[\widetilde{\MM^2}_{ijK}(x,y,z,t;p)]^4}\bigg\}
\eea
where various functions depending on the Feynman parameters are defined by
\bea\label{fu}
f_1^u (x,y,z,t)&=& \frac12 x(1-x)y^3z(1-z)t^3 (-1+3y +3tz)\nn\\
f_2^u(x,y,z,t) &=& x(1-x)y^4(1-y)z(1-z)t^4(1-t) \left[-1+2y+tz(1-y)(-3+5 t)\right]\nn\\
f_3^u(x,y,z,t) &=& 2x(1-x)y^4(1-y)^2z(1-z)t^4(1-t) [1+tz(3-4t)]\nn\\
f_4^u(x,y,z,t) &=& -3x(1-x)y^5(1-y)^3z^2(1-z)t^5(1-t)^3 \nn\\
f_5^u(x,y,z,t) &=& 6x(1-x)y^5(1-y)^3z^2(1-z)t^5(1-t)^3
\eea
for the up-like quarks, and
\bea
f_1^d (x,y,z,t)&=& \frac14 x(1-x)y^3z(1-z)t^3 [4-3y - 3(1-y)zt]\nn\\
f_2^d (x,y,z,t)&=& x(1-x)y^4(1-y)^2z(1-z)t^4(1-t)[1 - \frac12 z(1+t)] \nn\\
f_3^d (x,y,z,t) &=& x(1-x)y^4(1-y)z(1-z)t^4(1-t)[y + 2(1-y)z(\frac52 -2t)]\nn\\
f_4^d (x,y,z,t)&=& 0 \nn\\
f_5^d (x,y,z,t) &=& 3x(1-x)y^5(1-y)^3z^2(1-z)t^5(1-t)^3\nn
\eea
for the down-like quarks. We also introduced the shorthand notation
\bea\label{fd}
\MM^2_{ijK}(x,y,z,t;p) &:=& xyzt M_j^2 + (1-y)zt\me^2+y(1-y)(1-z)t M_W^2 \nn\\
&&+ y(1-y)(1-t)\md^2 - y(1-y)t(1-t)p^2\nn
\eea
and
\bea
\widetilde{\MM^2}_{ijK}(x,y,z,t;p) &:=& xyzt M_j^2 + (1-y)zt\me^2+y(1-y)(1-z)t M_W^2\nn\\
 &&+ y(1-y)(1-t)m_{u_K}^2 - y(1-y)t(1-t)p^2\nn
\eea
Notice the difference between up-like and down-like quarks in these expressions
(apart from the different masses of up- and down-like quarks):  although the
integrals (\ref{aqInt1}) and (\ref{aqInt2}) look almost the same, the  indices on
$q^\mu$, $\si^\rho$ and the loop momentum $k^\lambda$ are contracted differently

For small $p$ we arrive at
\bea
&& \!\!\!\!\!\!\!\!\!\!\!\!\!\!\!\!\!
 -\ri\cM_{auu}^{IJ}(p,q) = \\[1mm]
&=& q^\rho \bar\si_\rho \frac{g_2^2 }{128\sqrt{2}\pi^4\vev} \sum_{i,j,K} V^{IK}(V^\dag)^{KJ}|m^{ij}|^2M_j^2 \int_0^1 \rd x\int_0^1 \rd y\int_0^1 \rd z\int_0^1 \rd t \nn\\
&& \times \frac{x(1-x)y^3z(1-z)t^3 (-1+3y +3tz)}{[xyzt M_j^2 + (1-y)zt\me^2+y(1-y)(1-z)t M_W^2 + y(1-y)(1-t)m_{d_K}^2]^2} \nn
\eea
and
\bea
&&  \!\!\!\!\!\!\!\!\!\!\!\!\!\!\!\!\!
-\ri\cM_{add}^{IJ}(p,q) = \\[1mm]
&=& q^\rho \bar\si_\rho \frac{g_2^2 }{256\sqrt{2}\pi^4\vev} \sum_{i,j,K}
(V^\dag)^{IK}V^{KJ}|m^{ij}|^2M_j^2 \int_0^1 \rd x\int_0^1 \rd y\int_0^1 \rd z\int_0^1 \rd t \nn\\
&& \times \frac{x(1-x)y^3z(1-z)t^3 [4-3y - 3(1-y)zt]}{[xyzt M_j^2 + (1-y)zt\me^2+y(1-y)(1-z)t M_W^2 + y(1-y)(1-t)m_{u_K}^2]^2} \nn
\eea

In good approximation we can now put $m_{up} = m_{down} = 0$ (there are no IR divergences);
then the sum over $K$ can be performed, and by the unitarity of CKM matrix
the amplitudes become proportional to $\de^{IJ}$, {\it i.e.} flavor diagonal.
The non-degeneracy of the quark
masses makes possible quark flavor change in the interaction with the axion, but
the off-diagonal amplitudes are suppressed by factors of order
$\cO(m_{\text{quark}}^2/M_j^2)$. Note that the difference between the integrals
involving up and down quarks, respectively, is due not only to the CKM matrix and the different quark masses, but also to the different isospin and  the different topology of the diagrams; this leads to different formulae for the two cases.
Also, it appears that both the gauge invariant and non-invariant parts
of the $aWW$ amplitude are important, as both of them contribute to this amplitude,
as we can check in relations (\ref{gammarelations1}).

\section{Axion-gluon vertex}

After these preparations, we are ready at last to tackle the final part of the computation,
which will yield the coupling of $a(x)$ to gluons. In leading order, this coupling
is given by the set of three-loop diagrams depicted below. The first set consists of the following diagrams with insertions of the axion-quarks diagram determined before, with either up-quarks running in the loop

\begin{center}
\includegraphics[width=11cm,viewport= 150 620 480 720,clip]{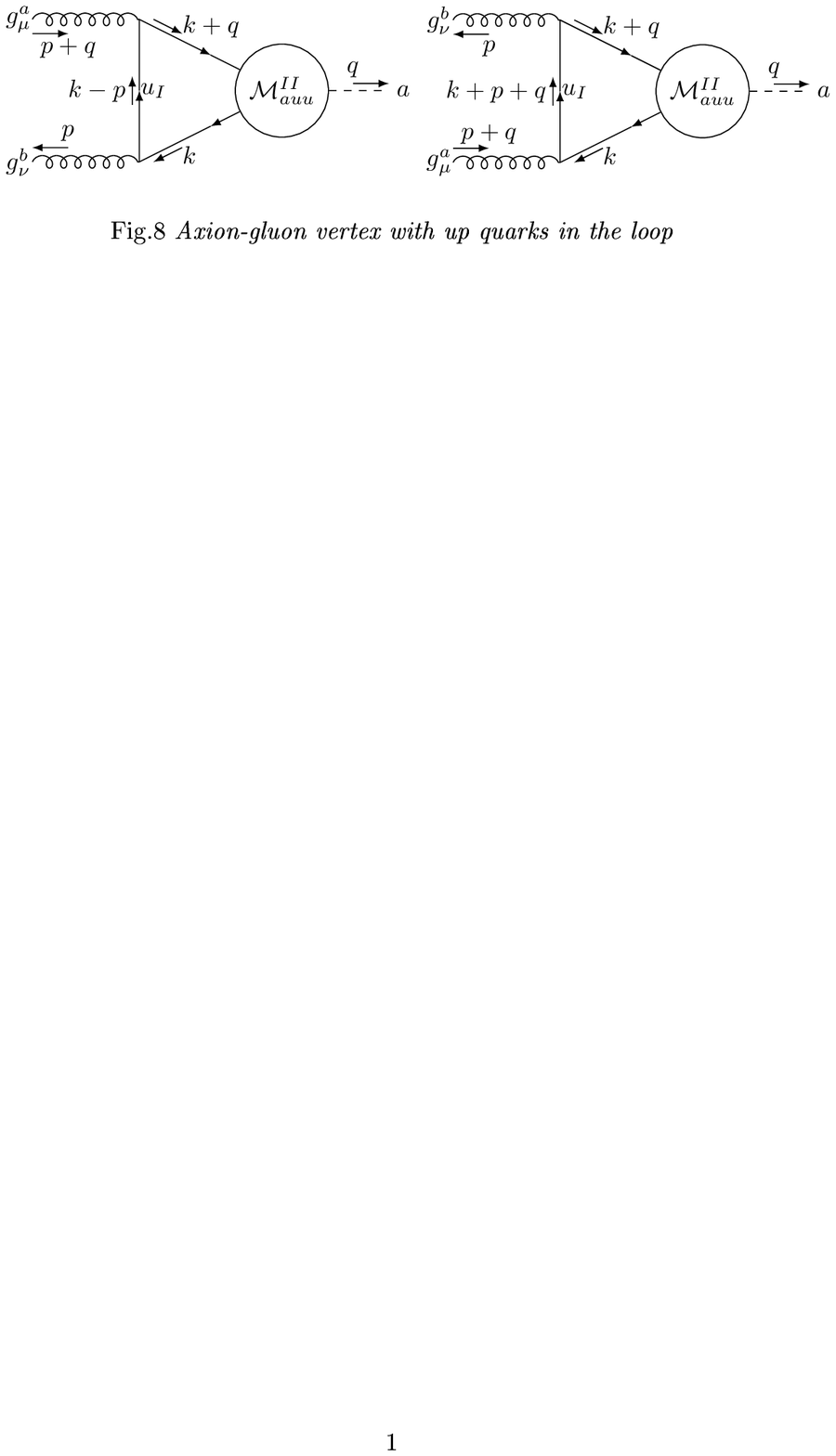} \end{center}

or with down-quarks:

\begin{center}
\includegraphics[width=11cm,viewport= 150 620 480 720,clip]{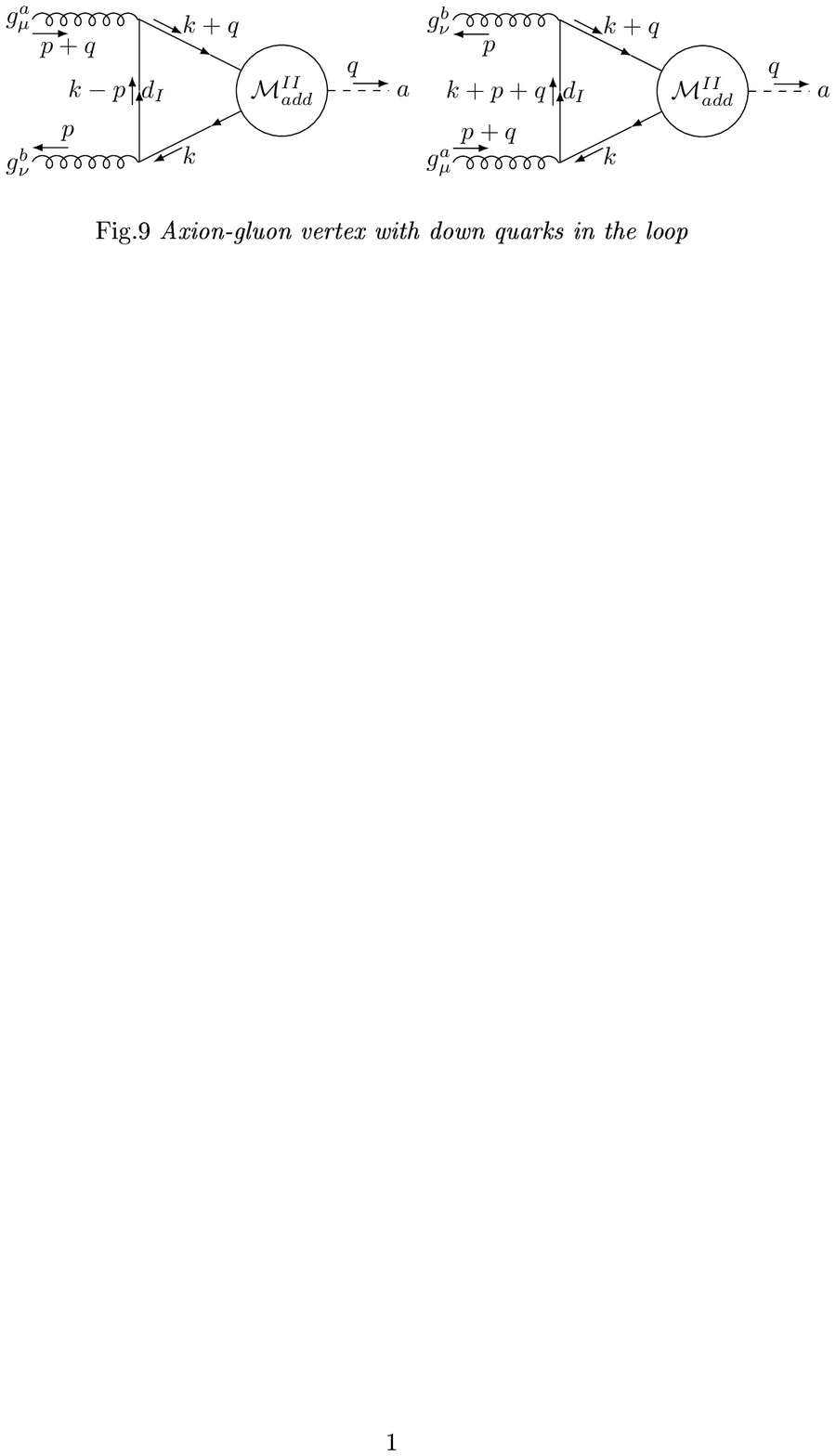} \end{center}

However, there are also the following non-planar diagrams:

\begin{center}
\includegraphics[width=8cm,viewport= 150 510 400 720,clip]{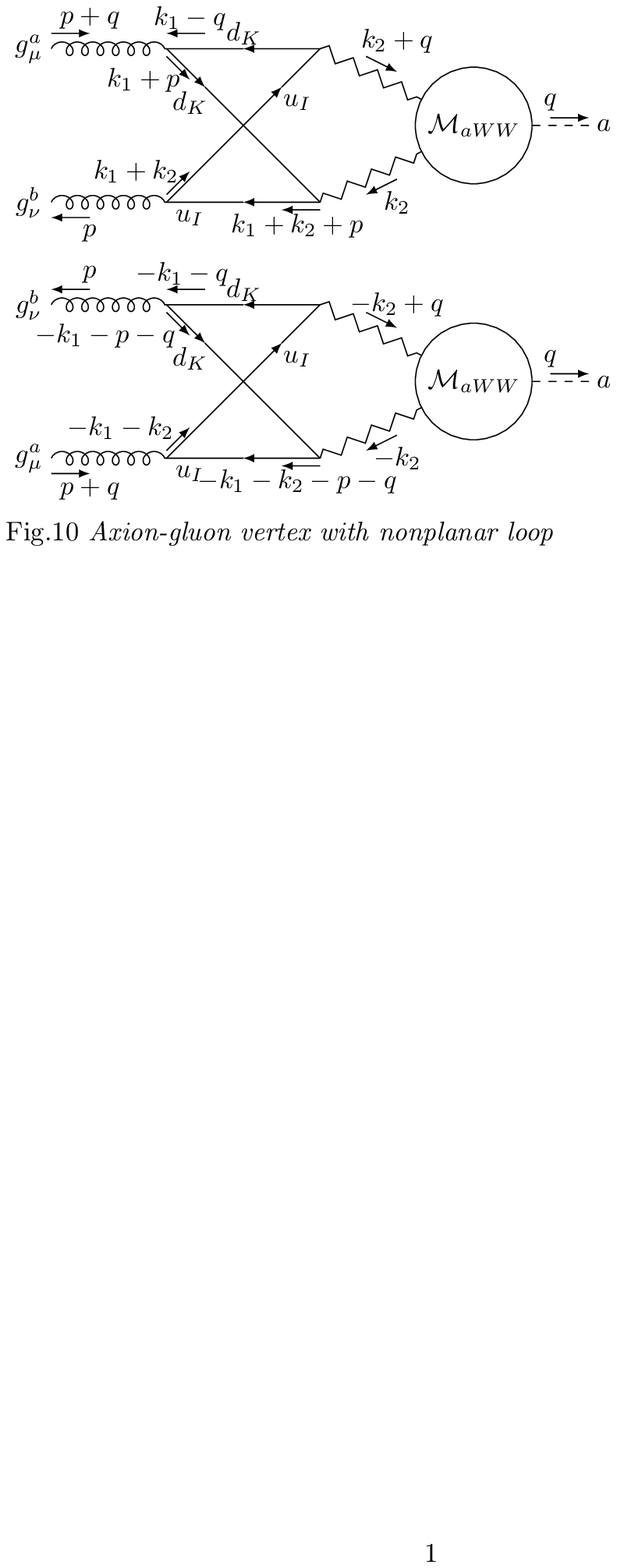} \end{center}

We first of all see that the part of the total (summed) amplitude linear in $q$ is
antisymmetric under the simultaneous exchange $\mu\lrarr\nu$, $p\rarr-p$. This
means that the only possible tensor structures in the amplitude are either proportional
to $p^\mu q^\nu-p^\nu q^\mu$ or to $\eps^{\mu\nu\la\si}p_\la q_\si$. As we already
explained, the first structure can be excluded by gauge invariance, which leaves only the second contribution. Of course, this claim is confirmed by the explicit calculation.
Hence, the effective interaction of axion and gluons for sufficiently small $q$
is indeed of the form (\ref{Lagg}), with a non-vanishing (but small) coefficient.

The full integrals are now very cumbersome, so we try to present the result in a
compact form. As already mentioned in the introduction, we revert to 4-spinor notation for the loops {\em not} involving neutrino lines. For the first two diagrams we have, to leading order in $q$,
\bea
&&  \!\!\!\!\!\!\!\!\!\!\!\!\!\!\!\!\!
-\ri \cM^{ab \mu\nu}_{(agg)} (\text{through up quarks}) = \nn\\[1mm]
&=& -\ri\de^{ab}\,\frac{g_3^2}{2}\, \sum_I \int \frac{d^4k}{(2\pi)^4} \times\nn \\
&& \times \Bigg[\text{Tr}\left\{\frac{\kslash+\qslash+m_{u_I}}{(k+q)^2-m_{u_I}^2} \ga^\mu \frac{\kslash-\pslash+m_{u_I}}{(k-p)^2-m_{u_I}^2} \ga^\nu \frac{\kslash+m_{u_I}}{k^2-m_{u_I}^2}
\big[-\ri\cM_{auu}^{II}(k,q)\big]\right\}  \nn \\
&& \qquad\qquad + \; \big(\mu \lrarr \nu\,,\, p\rarr -p-q\big)\Bigg]  = \nn \\
&=& -\ri q^\rho \de^{ab}\frac{g_3^2g_2^4}{128\sqrt{2}\pi^4\vev}\sum_{i,j,I,K} |V^{IK}|^2|m^{ij}|^2M_j^2 \times\nn\\
&&
\times\int_0^1 \rd x\int_0^1 \rd y\int_0^1 \rd z\int_0^1 \rd t \int \frac{d^4k}{(2\pi)^4} \times\nn \\
&& \left[\text{Tr}\left\{\frac{\kslash+m_{u_I}}{k^2-m_{u_I}^2} \ga^\mu \frac{\kslash-\pslash+m_{u_I}}{(k-p)^2-m_{u_I}^2} \ga^\nu \frac{\kslash+m_{u_I}}{k^2-m_{u_I}^2} \ga^\si P_L \right\} + (\mu \lrarr \nu, p\rarr -p)\right] \nn\\
&& \times \bigg\{\frac{f_1^u(x,y,z,t) \eta_{\rho\si}}{[\MM^2_{ijK}(x,y,z,t;k)]^2} + \frac{f_2^u(x,y,z,t)k^2 \eta_{\rho\si} + f_3^u(x,y,z,t) k_\rho k_\si}{[\MM^2_{ijK}(x,y,z,t;k)]^3}  \nn\\
&&\qquad\qquad\qquad
+\;\frac{f_4^u(x,y,z,t)k^4\eta_{\rho\si} + f_5^u(x,y,z,t)k^2 k_\rho k_\si}{[\MM^2_{ijK}(x,y,z,t;k)]^4}
\bigg\}
\eea
where the expression for $\cM^{IJ}_{auu}(k,q)$ obtained in the foregoing section
has been used. The remaining momentum  space integral can
be done in the standard way, introducing yet another Feynman
parameter $u$ (so there are now altogether {\em five} Feynman parameters $x,y,z,t,u$).
Packaging the previous results into various new functions we get,
after some calculations,
\bea && \!\!\!\!\!\!\!\!\!\!\!\!\!\!\!\!\! \!\!\!\!\!\!\!\!\!\!\!\!
-\ri \cM^{ab \mu\nu}_{(agg)} (\text{through up quarks}) = \nn\\[1mm]
&=&  \ri\eps^{\mu\nu\xi\si}p_\xi q_\si \de^{ab}  \frac{g_3^2g_2^4}{32\sqrt{2}(2\pi)^6\vev} \sum_{i,j,I,K} |V^{IK}|^2|m^{ij}|^2M_j^2  \times \\
&& \times \int_0^1 \rd x\int_0^1 \rd y\int_0^1 \rd z\int_0^1 \rd t\int_0^1 \rd u \frac{F_1(x,y,z,t,u)}{[\cM^2_{ijKI}]^2}+ O(p^2)\nn
\eea
where
\bea
\cM^2_{ijKI} &=& xyztu\, M_j^2 + (1-y)ztu\,\me^2+y(1-y)(1-z)tu\, M_W^2\nn\\
 &&+ y(1-y)(1-t)u\,\md^2 + y(1-y)t(1-t)(1-u)m_{u_I}^2
\eea
and the function
\bea
F_1(x,y,z,t,u) &=& u(1-u)\left[(-4u)f_1^u+[(-2+6u) f_2^u-f_3^u]\frac{u}{y(1-y)t(1-t)}\right. \nn\\
&& \qquad\qquad \left.+\;  \big[(4-8u)f_4^u-f_5^u\big]\frac{u^2}{[y(1-y)t(1-t)]^2}\right]
\eea
itself depends on the previous Feynman parameter functions (\ref{fu}) and (\ref{fd}),
but now with an extra dependence on the fifth Feynman parameter $u$.

Similarly, the third and fourth diagrams give
\bea
&&  \!\!\!\!\!\!\!\!\!\!\!\!\!\!\!\!\! \!\!\!\!\!\!\!\!\!\!\!\!
-\ri \cM^{ab \mu\nu}_{(agg)} (\text{through down quarks}) = \nn\\[1mm]
&=&  \ri\eps^{\mu\nu\xi\si}p_\xi q_\si \de^{ab}  \frac{g_3^2g_2^4}{32\sqrt{2}(2\pi)^6\vev} \sum_{i,j,I,K} |V^{IK}|^2|m^{ij}|^2M_j^2  \times \\
&& \times \int_0^1 \rd x\int_0^1 \rd y\int_0^1 \rd z\int_0^1 \rd t\int_0^1 \rd u \frac{F_2(x,y,z,t,u)}{[\tilde \cM^2_{ijKI}]^2}+ O(p^2)\nn
\eea
Here we have subsumed the previous results into the functions
\bea
\tilde \cM^2_{ijKI} &=& xyztu\, M_j^2 + (1-y)ztu\,\me^2+
y(1-y)(1-z)tu\, M_W^2 \nn\\
&&+ y(1-y)(1-t)u\,m_{u_I}^2 + y(1-y)t(1-t)(1-u)\md^2
\eea
and
\bea
F_2(x,y,z,t,u) &=& u(1-u)\left[(-4u)f_1^d+[(-2+6u) f_2^d-f_3^d]\frac{u}{y(1-y)t(1-t)}\right. \nn\\
&& \qquad \left.+ \big[(4-8u)f_4^d-f_5^d\big]\frac{u^2}{[y(1-y)t(1-t)]^2}\right]
\eea
Finally, the last two diagrams must be computed directly. To leading order in $q$,
they are given by formulae
\bea
&& -\ri \cM^{ab \mu\nu}_{(agg)} (\text{nonplanar diagrams}) =  \nn\\[1mm]
&=& -\de^{ab}\frac{g_3^2g_2^2}{4}\sum_{I,K} |V^{IK}|^2 \int \frac{d^4k_1}{(2\pi)^4} \int \frac{d^4k_2}{(2\pi)^4} \left(-\ri\cM_{aWW}^{\ka\la}(k_2,q)\right) \left(\frac{1}{k_2^2-M_W^2}\right)^2 \nn \\
&& \times \frac{1}{(k_1+p)^2-m_{d_K}^2}\frac{1}{k_1^2-m_{d_K}^2}
\frac{1}{(k_1+k_2)^2-m_{u_I}^2}\frac{1}{(k_1+k_2+p)^2-m_{u_I}^2}  \times \nn \\
&& \times \text{Tr}\bigg\{(\kslash_1+\pslash+m_{d_K})\ga^\mu(\kslash_1+m_{d_K})\ga_\ka P_L \times \nn \\
&& \hspace{2cm} \times (\kslash_1+\kslash_2+m_{u_I})\ga^\nu (\kslash_1+\kslash_2+\pslash+m_{u_I})\ga_\la P_L\bigg\}  \nn
\eea
\bea
&& -\de^{ab}\frac{g_3^2g_2^2}{4}\sum_{I,K} |V^{IK}|^2 \int \frac{d^4k_1}{(2\pi)^4} \int \frac{d^4k_2}{(2\pi)^4} \left(-\ri\cM_{aWW}^{\ka\la}(-k_2,q)\right) \left(\frac{1}{k_2^2-M_W^2}\right)^2 \nn\\
&& \times \frac{1}{(k_1+p)^2-m_{d_K}^2}\frac{1}{k_1^2-m_{d_K}^2}
\frac{1}{(k_1+k_2)^2-m_{u_I}^2}\frac{1}{(k_1+k_2+p)^2-m_{u_I}^2}  \times \nn \\
&& \times \text{Tr}\bigg\{(-\kslash_1-\pslash+m_{d_K})\ga^\nu(-\kslash_1+m_{d_K})\ga_\ka P_L  \times \nn \\
&& \hspace{2cm} \times (-\kslash_1-\kslash_2+m_{u_I})\ga^\mu (-\kslash_1-\kslash_2-\pslash+m_{u_I})\ga_\la P_L\bigg\}  = \nn\\
&=& -\de^{ab}\frac{g_3^2g_2^2}{4}\sum_{I,K} |V^{IK}|^2 \int \frac{d^4k_1}{(2\pi)^4} \int \frac{d^4k_2}{(2\pi)^4} \left(\frac{1}{k_2^2-M_W^2}\right)^2 \times \nn \\
&& \times \frac{1}{(k_1+p)^2-m_{d_K}^2}\frac{1}{k_1^2-m_{d_K}^2}
\frac{1}{(k_1+k_2)^2-m_{u_I}^2}\frac{1}{(k_1+k_2+p)^2-m_{u_I}^2}  \times \nn \\[1mm]
&&\qquad  \times \; T^{\mu\nu}_{\ka\la}(k_1,k_2,p)\left[-\ri\cM_{aWW}^{\ka\la}(k_2,q)\right]
\eea
where
\bea
T^{\mu\nu}_{\ka\la}(k_1,k_2,p) &:=&  (k_1+p)^\al(k_1+k_2+p)^\be \times  \\[1mm]
&& \times \big[\, \text{Tr}\left\{\ga_\al\ga^\mu\kslash_1\ga_\ka (\kslash_1+\kslash_2)\ga^\nu \ga_\be\ga_\la P_L\right\}- (\mu \leftrightarrow \nu) \big]  \nn \\[1mm]
&& + \, m_{u_I}^2 (k+p)^\al \big[\,\text{Tr}\left\{\ga_\al\ga^\mu\kslash_1\ga_\ka \ga^\nu\ga_\la P_L\right\} - (\mu \leftrightarrow \nu)\big] + \cO(m_{d_K}^2)  \nonumber
\eea
After a tedious calculation we obtain the result
\bea
&& \!\!\!\!\!\!\!\!\!\!\!\!\!\!\!\!\!\!\!\!\!\!\!\!\!\!\!
-\ri\cM^{ab\mu\nu}_{agg}\text{(non-planar diagrams)} = \nn\\
&=& \ri\eps^{\mu\nu\rho\si}p_\rho q_\si \de^{ab}\frac{g_3^2g_2^4}{32\sqrt{2}(2\pi)^6\vev}\,\sum_{I,K,i,j} |V^{IK}|^2|m^{ij}|^2M_j^2 \times \\
&& \qquad \qquad \times \int_0^1 \rd x\int_0^1 \rd y\int_0^1 \rd z\int_0^1 \rd t\int_0^1 \rd u\frac{F_3(x,y,z,t,u)}{[\tilde \cM^2_{ijKI}]^2}\nn
\eea
where
 \bea
&&F_3(x,y,z,t,u) = 3x(1-x)y^3z(1-z)t^2u^2(1-u)\left\{2t^2u(9-4u)(1-y)z\right.\nn\\
&&\left.\quad\quad - t[z(1-y)(-8u^2+15u+1)+2(1-u)(1-2y)] -u(1-y)(z+2)\right\}\nn
\eea
In total, we thus arrive at the final result equivalent to (\ref{Lagg})
\beq
-\ri\cM^{ab\mu\nu}_{agg}\text{(total)} = -\ri\frac{g_3^2}{16\pi^2 f_a}\eps^{\mu\nu\rho\si}
p_\rho q_\si \de^{ab}
\eeq
where the axion coupling is given by
\bea\label{faxion}
f_a^{-1}  &=& - \frac{g_2^4}{128\sqrt{2}\pi^4\vev}\,\sum_{I,K,i,j} |V^{IK}|^2|m^{ij}|^2M_j^2 \times  \\
&& \qquad  \times \int_0^1 \rd x\int_0^1 \rd y\int_0^1 \rd z\int_0^1 \rd t\int_0^1 \rd u \left(\frac{F_1}{[\cM^2_{ijKI}]^2}+\frac{F_2+F_3}{[\tilde \cM^2_{ijKI}]^2}\right)  \nn
\eea
This integral cannot be evaluated in closed form, but we can easily get a numerical
estimate. First, to recover the order of magnitude estimate (\ref{faN}) quoted in the introduction, we take the heavy neutrino masses degenerate, that is,  $M_j = M$, so we can exploit the unitarity relation $\sum_K  |V^{IK}|^2 = 1$. Then using
$\al_w =g_2^2/(4\pi)$ and $\sum_{i,j}|m^{ij}|^2 \approx \vev \sum m_{\nu}$,
we see that the remaining integral is of order $\cM^{-2}$ where $\cM$ is the larger
of the two values $M$ and $M_W$. For the actual numerical evaluation we can also
neglect the quark and lepton masses. Setting $M_W=80.4 \text{ GeV}$,
$m_{top} = 172.9 \text{ GeV}$, $\sum m_{\nu} = 1\text{ eV}$, and (as an example)
$\vev = 400\text{ GeV}$ we get for various values of $M$ the following numbers

\begin{center}\begin{tabular} {c | c}
M [GeV] & $f_a [10^{18}\text{ GeV}]$ \\
\hline
100 & 3.1 \\
150 & 2.6 \\
300 & 2.2 \\
500 & 2.1 \\
700 & 2.2 \\
1000 & 2.3
\end{tabular}\end{center}

As we already pointed out in the introduction, these values may be affected by
higher order QCD corrections, because $\al_s$ is large at small momenta.

\vspace{5mm}

\noindent
{\bf Acknowledgment:} We are grateful to Wilfried Buchm\"uller for enlightening comments
on a first version of this paper. H.N. would also like to thank Manfred Lindner for discussions.

\newpage
\appendix
\section{Weyl-spinor conventions}
Since we heavily use two-spinor notation throughout this paper we here briefly summarize our conventions and notations, see \cite{BW} for more information.
Employing the `mostly minus' metric $\eta_{\mu\nu} = \text{diag}(+1,-1,-1,-1)$
we define
$$
\si^\mu_{\al\dal} = ({\bf{1}}\,,\,\si^i)\; , \quad
\bsi^{\mu \dal\al} = \eps^{\dal\dbe} \eps^{\al\be} \si^\mu_{\be\dbe}
\equiv ({\bf{1}}\,,\, - \si^i)
$$
where $\eps^{12}=\eps_{21} = -\eps_{12} = -\eps^{21} = 1$ and $\eps_{11}=\eps_{22} =0$,
hence $\eps_{\al\ga} \eps^{\ga\be} = \de_\al^\be$ (the definitions for $\eps^{\dal\dbe}$
are the same). Then we have
$$
\si^\mu \bsi^\nu + \si^\nu\bsi^\mu = 2\,\eta^{\mu\nu}
$$
as well as the completeness relations
$$
{\rm Tr}\, \si^\mu \bsi^\nu = 2\,\eta^{\mu\nu}\;, \quad
\si^\mu_{\al\dal} \, \bsi_\mu^{\dbe\be} = 2 \, \delta_\al^\be \delta_{\dot\alpha}^{\dot\beta}
$$
Furthermore
$$
{\rm Tr}\, \si^\mu \bsi^\nu \si^\rho\bsi^\lambda =
2\big(\eta^{\mu\nu}\eta^{\rho\lambda} - \eta^{\mu\rho}\eta^{\nu\lambda}
     + \eta^{\mu\lambda} \eta^{\nu\rho} - \ri\eps^{\mu\nu\rho\lambda}\big)
$$
where $\eps^{0123} = 1$. To relate 2-spinors to 4-spinors we need the
Dirac $\ga$-matrices and the charge conjugation matrix $\cC$, which are
given by, respectively,
$$
\ga^\mu =
 \left( \begin{array}{cc} 0 & \si^\mu_{\al\dbe} \\  \bsi^{\mu\dal\be} & 0 \end{array}\right) \;, \;\;
\ga^5 = \ri \ga^0\ga^1\ga^2\ga^3 =
 \left( \begin{array}{cc} \delta_\alpha^\beta & 0 \\  0 & - \,\delta^{\dal}_{\dbe} \end{array}\right) \;,\;\;
\cC =  \left( \begin{array}{cc} \eps_{\al\be} & 0 \\  0 & \eps^{\dal\dbe} \end{array}\right) \;.
$$
A Dirac 4-spinor $\Psi$ then decomposes into two Weyl spinors via
$$
\Psi \equiv \left(\begin{array}{c} \vp_\alpha\\[1mm]
                                                         \bar\chi^{\dot\alpha} \end{array}\right)
\equiv \left(\begin{array}{c} \Psi_{L \alpha}\\[1mm]
                                                         \bar\Psi_R^{\dot\alpha} \end{array}\right)
\;\;\Rightarrow\;\;
\overline{\Psi} \equiv \Psi^\dagger\ga^0  = (\chi^\al , \bar\vp_{\dal})
$$
where the indices on Weyl spinors are always pulled up and down
{\em from the left}, e.g. $\vp^\al = \eps^{\al\be} \vp_\be \,\Rightarrow
\vp_\al = \eps_{\al\be} \vp^\be$; furthermore $(\vp_\al)^\dagger = \bar\vp_{\dal}$,
{\it etc.} The charge conjugate spinor is
$$
\Psi^c \equiv \cC \overline{\Psi}^T =
\left(\begin{array}{c} \chi_\alpha\\[1mm]
                                                         \bar\vp^{\dot\alpha} \end{array}\right)
$$
so $\Psi$ is Majorana if $\vp_\al = \chi_\al$. Note that in the main text we label the independent Weyl spinors by the subscripts $L$ and $R$, as in (\ref{Qi}) and (\ref{ui}), with the exception of the neutrino 4-spinor, for which we use different letters $\nu$ and $N$.
This is done mainly in order not to encumber the notation with too many different letters, although it does not quite conform to standard usage, where $\Psi_{L,R}$ are usually defined as the degenerate (projected) 4-component spinors $\frac12 (1 \mp \ga^5)\Psi$. Finally, we recall that hermitian conjugation
inverts the position of the (anti-commuting) fermionic operators, {\it i.e.}
$(\vp\chi \cdots \psi)^\dagger = \psi^\dagger\cdots \chi^\dagger\vp^\dagger$.

\end{document}